\documentclass[a4paper,UKenglish]{lipics-v2016}
\pdfoutput=1

\usepackage[utf8]{inputenc}
\usepackage{mathtools}
\usepackage{multicol}
\usepackage{multirow}
\usepackage{booktabs}
\usepackage{courier}
\usepackage[scaled]{helvet}
\usepackage{url}
\usepackage{listings}
\usepackage{enumitem}
\usepackage{graphicx}
\usepackage[table]{xcolor}
\usepackage{amsmath}
\usepackage{mdwlist}
\usepackage{pifont}
\usepackage[export]{adjustbox} 
\usepackage{wrapfig}

\usepackage[firstpage]{draftwatermark}
\SetWatermarkText{\hspace*{6.5in}\raisebox{8in}{\includegraphics[scale=0.1]{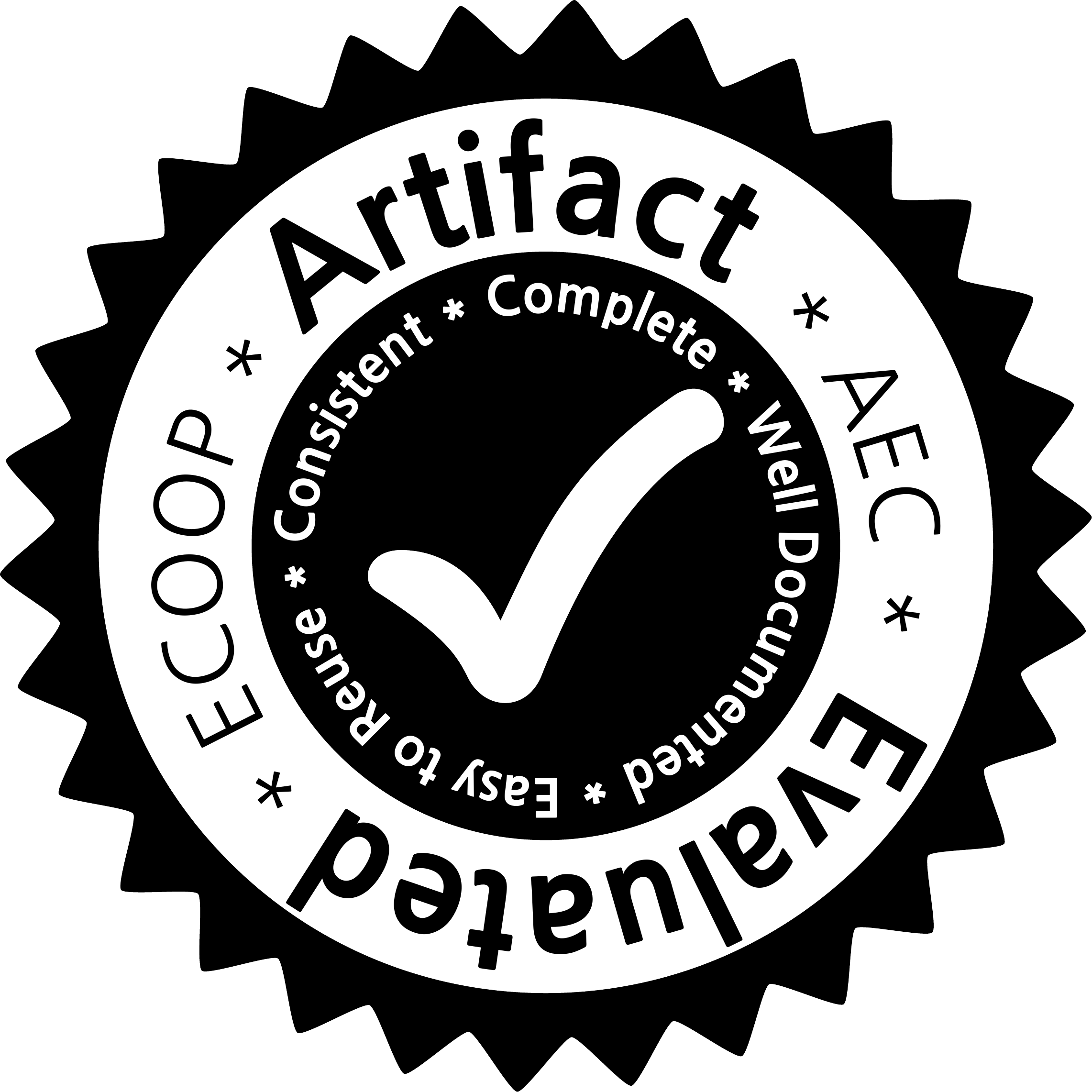}}}
\SetWatermarkAngle{0}

\usepackage{xspace}
\newcommand{\ourvm}{PyHyp\xspace}
\newcommand{\hippy}{HippyVM\xspace}
\newcommand{\pypy}{PyPy\xspace}

\newcommand{\ourvmcomp}{PyHyp$_{\textrm{PHP}}$\xspace}
\newcommand{\ourvmcompr}{PyHyp$_{\textrm{Py}}$\xspace}
\newcommand{\ourvmmono}{PyHyp$_{\textrm{mono}}$\xspace}

\EventEditors{John Q. Open and Joan R. Acces}
\EventNoEds{2}
\EventLongTitle{42nd Conference on Very Important Topics (CVIT 2016)}
\EventShortTitle{CVIT 2016}
\EventAcronym{CVIT}
\EventYear{2016}
\EventDate{December 24--27, 2016}
\EventLocation{Little Whinging, United Kingdom}
\EventLogo{}
\SeriesVolume{42}
\ArticleNo{23}

\begin{document}


\title{Fine-grained Language Composition:\newline A Case Study}
\titlerunning{Fine-grained Language Composition: A Case Study}
\author{\href{http://eddbarret.co.uk/}{Edd Barrett}}
\author{\href{http://cfbolz.de/}{Carl Friedrich Bolz}}
\author{\href{http://lukasdiekmann.com/}{Lukas Diekmann}}
\author{\href{http://tratt.net/laurie/}{Laurence Tratt}}
\affil{Software Development Team, Department of Informatics, King's College London. \href{http://soft-dev.org/}{http://soft-dev.org/}}

\Copyright{Edd Barrett, Carl Friedrich Bolz, Lukas Diekmann and Laurence Tratt}


\maketitle


\begin{abstract}
Although run-time language composition is common, it normally takes the form of
a crude Foreign Function Interface (FFI). While useful, such compositions
tend to be coarse-grained and slow. In this paper we introduce a novel fine-grained
syntactic composition of PHP and Python which allows users to embed each
language inside the other, including referencing variables across languages.
This composition raises novel design and implementation challenges. We show
that good solutions can be found to the design challenges; and that the
resulting implementation imposes an acceptable
performance overhead of, at most, 2.6x.
\end{abstract}

\section{Introduction}
\label{sec:intro}

\begin{figure*}[t]
\centering
\includegraphics[width=.9\textwidth]{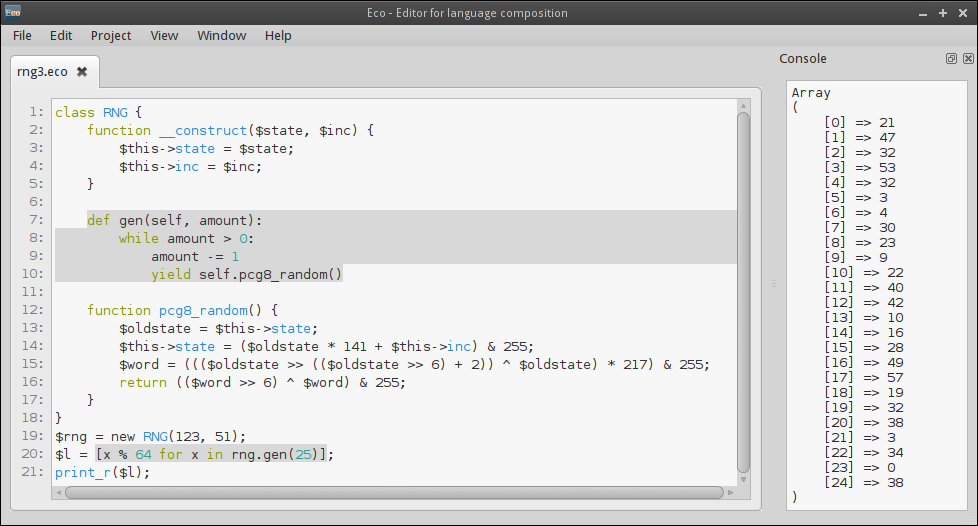}
\begin{picture}(0,0)
\put(-230, 102){\textcolor{black}{\ding{182}}}
\put(-235, 24){\textcolor{black}{\ding{183}}}
\put(-24, 100.5){\textcolor{black}{\ding{184}}}
\end{picture}
\caption{A PCG8 pseudo-random number generator~\cite{oneill15pcg} \ourvm program, written
in the Eco language composition editor. In this case, the composed PHP and Python
program will be exported to
\ourvm compatible source code. The outer (white background) parts of the file are written in PHP,
the inner (grey background) parts of the file in Python. \ding{182} A Python
language box is used to add a
generator method written in Python to the PHP class \texttt{RNG}.
\ding{183} A Python language box is used to embed a Python expression
inside PHP, including a cross-language variable reference for \texttt{rng}
(defined in line 19 in PHP and referenced in line 20 in Python). In
this case, a Python list comprehension builds a
list of random numbers. When the list is passed to PHP, it is
`adapted' as a PHP array. \ding{184} Running the program pretty-prints the
adapted Python list as a PHP array.}
\label{fig:rngexample}
\end{figure*}

Language composition allows programmers to create systems written in a mix of
programming languages. Most commonly,
a Foreign Function Interface (FFI) to C is provided so that programs can interact with external libraries.
However, other instances of language composition are rare, as crossing the
barrier between arbitrary languages is difficult. In many cases, the only way
to do so is by having different languages run their parts of the system
in separate processes that communicate using (slow) inter-process communication
mechanisms. The most common alternative is to use a single Virtual Machine (VM)
(e.g.~a Java VM), and translate all languages into that VM's bytecode format.
This enables finer-grained compositions, but their performance is still
generally underwhelming~\cite{barrett15approaches}.

We believe that there are two different types of friction which make good
language compositions difficult: \emph{semantic friction}
occurs when an aspect of one language has no equivalent in the other;
and \emph{performance friction} occurs when the implementation of one language's
behaviour forces the other to execute slowly.

Our hypothesis is that it is possible to reduce the currently accepted levels of
friction in language compositions. We believe that the only way to test this
hypothesis is through a concrete case study, since friction manifests in different
ways in each language composition. We therefore composed together two
real-world, widely used languages, Python and PHP, to make a new language
composition called \ourvm. At a low-level, \ourvm defines a
(somewhat traditional) FFI between PHP and Python that allows cross-language
calls and the exchange of data. Building on the FFI, \ourvm then provides the
basis for a novel syntactic composition. As shown in
Figure~\ref{fig:rngexample}, a single file can contain multiple fragments of PHP
and Python code, and variables can be
referenced across different language fragments (e.g. Python code can `see' PHP
variables and vice versa). Unlike approaches which translate
one language into another, \ourvm does not alter existing
language semantics, nor does it limit users to a subset of either language.

Depending on how one chooses to classify programming languages, Python and PHP
can appear similar---most obviously, both are dynamically typed. From our
perspective, however, there are a number of tricky differences: PHP has multiple
global namespaces which can span multiple files, whereas Python uses one global
namespace per file (semantic friction); most of PHP's core data-structures are
immutable, whereas many of Python's are mutable (semantic and performance friction); and
PHP's sole collection data type is a mapping, whereas Python separates the
notion of mappings from that of sequences (semantic and performance friction).
As this may suggest, this combination of languages presents a number of
design and implementation challenges which have no obvious precedent.
We show that \ourvm's solutions to these challenges allow
interesting case studies to be implemented.

The practicality of our work rests on two recent developments.
First, and most important, is the concept of interpreter
composition. The basic idea is to make use of systems which can
generate Just-In-Time (JIT) compiled VMs solely from the description of an interpreter
(e.g.~RPython~\cite{bolz14impact} or Truffle~\cite{wuerthinger13onevm}). There
are three existing compositions in this style: Prolog and Python~\cite{barrett15approaches};
C and Ruby~\cite{grimmer2014dynamically}; and C, Ruby, and JavaScript~\cite{grimmer15interop}.
In essence, each of these systems implements a
traditional FFI between its constituent interpreters. All of the systems have good peak
performance, but implement simple FFIs. \ourvm
defines a much finer-grained FFI between its two languages and also enables
syntactic composition between the two languages. The second concept we make use of is
language boxes as found in the Eco
editor~\cite{diekmann14eco}, which allow users to naturally write fragments
of different languages alongside each other. Note that \ourvm neither
extends, nor requires, Eco; however, Eco does hide several tedious details
from users.

Although universal answers to our hypothesis are impossible, \ourvm shows
that it is possible to create compositions which validate our
hypothesis. To summarise, we show that:

\vspace{-5pt}
\begin{enumerate*}
    \item \ourvm's FFI addresses a number of challenging semantic
        friction points.
    \item Syntactic composition is possible and that practical
        designs can be found for novel cross-language features.
    \item \ourvm's fine-grained language composition has, in the worst case,
        a performance overhead of 2.6x over its mono-language constituents.
\end{enumerate*}
\vspace{-5pt}

\noindent A VirtualBox VM containing repeatable experiments, data and case
studies is available at:%
\vspace{-0.2em}
\begin{center}
\small
\url{https://archive.org/download/ecoop16_pyhyp_artefacts/v0.2/ecoop16_pyhyp_vm_0.2.tar.gz}
\end{center}

\section{Background}

We assume a basic knowledge of Python syntax and semantics, but not of
meta-tracing, interpreter composition, or PHP. In this section we provide
overviews of the latter three.

\subsection{Meta-tracing}
\label{sec:metatrace}

Tracing JIT compilers record hot loops (`traces') in an interpreted program,
optimise those traces, and then compile them into machine
code~\cite{bala00dynamo,gal06hotpathvm}. An individual trace is thus a record of
one particular path through a program's control flow graph. Subsequent
executions which follow that same path can use the machine code generated from
the trace instead of the (slow) interpreter. To ensure that the path followed
really is the same, `guards' are left in the machine code at every possible
point of divergence. If a guard fails, execution then reverts back to the
interpreter.

Meta-tracing JITs have the same basic model, but replace the manually written
tracer and machine code generator with equivalents automatically generated
from the interpreter itself~\cite{mitchell70design,sullivan03dynamic,yermolovich09optimization,bebenita10spur,bolz14impact}. The key to good meta-tracing performance is heavily optimised traces. Language
implementers can annotate the interpreter provide `hints' to the meta-tracer to improve the quality of
compiled traces. For example, hints can be used to mark parts of the
interpreter constant, allowing the trace optimiser to apply constant folding.
\label{elidable functions} Similarly, hints can
mark a function in the interpreter as `elidable': given the
same inputs, it always returns the same outputs.\footnote{Unlike pure functions,
elidable functions can have idempotent side-effects (e.g.~caching). The user
is responsible for guaranteeing that the relationship between inputs and outputs is maintained.}
The optimiser can then
replace calls to (slow) elidable functions with (fast) checks on input
values, substituting the output values in place of the function call.
Identifying parts of an interpreter amenable to such hints requires the author's
knowledge of both the semantics of the language being implemented and common
idioms of use.

In this paper we use RPython, the main extant meta-tracing language. RPython
is a statically typed subset of Python with a type system similar to Java's.
Unlike seemingly similar languages
(e.g.~Slang~\cite{ingalls97back} or PreScheme~\cite{kelsey94tractable}), RPython
is more than just a thin layer over C:~it is, for example, fully garbage
collected and has several high-level data types (e.g.~lists and dictionaries).
Despite this, VMs written in RPython have performance levels which far exceed
traditional interpreters~\cite{bolz14impact}.
The specific details
of RPython are generally unimportant in most of this paper, and we do not
dwell on them: we believe that one could substitute any reasonable
meta-tracing language (or its cousin approach, self-optimising interpreters with dynamic partial
evaluation~\cite{wuerthinger13onevm}) and achieve similar results.

\subsection{Interpreter Composition}

Interpreter composition involves `glueing' together two or more existing
interpreters such that each can utilise the other. Assuming that both
interpreters are written in the same language, a basic composition
is simple: interpreter $A$ needs to import interpreter $B$ and then call
appropriate functionality in $B$. A more sophisticated composition will define
matters such as data type conversion, and how and when execution passes between
its constituent interpreters. Achieving the desired composition can require
adding entirely new glue code,
and/or invasively modifying the constituent interpreters.

Since, traditionally, interpreters are slow, composing them tends to worsen
performance~\cite{barrett15approaches}. Fortunately, composed interpreters are as amenable to
meta-tracing -- and its close cousin dynamic partial evaluation~\cite{wuerthinger13onevm} --
as their constituent interpreters. Existing
examples of such compositions include Python and Prolog~\cite{barrett15approaches}
and Ruby, C and JavaScript~\cite{grimmer15interop}. Both systems have good
peak performance.

In the rest of this paper, we use `interpreter' to mean the source-code of the
system that is used to produce an executable binary `VM'. In practice, most readers
can consider the terms `interpreter' and `VM' to be interchangeable with only
a small loss of precision.

\subsection{PHP}

PHP is a language used widely for server-side webpage creation. Originally intended as a language
to glue together CGI programs written in C, it
gradually evolved into a complete programming language. Largely due to this
gradual evolution, PHP features a number of design decisions that appear unusual
when viewed from the perspective of other imperative languages such as Python or
Ruby. As one example that appears later in the paper shows, many of PHP's primitive
data types, including arrays, are immutable.
We give further details on such features as needed.

PHP's syntax is Perl-esque, taking influence from the Unix shell
(e.g.~variables start with a `\$') and C (e.g.~basic control structures).
The following (contrived) example shows most of the syntax needed to understand
this paper's examples:
\begin{lstlisting}[label=lst:phpoverview, numbers=left]
$i=1;                 // Assign 1 to variable $i
$j=&$i;               // Create a reference to var $i
$a=array(3, 4, 5);    // Create a list-like array
$b=array("bob"=>45);  // Create a dictionary-like array
foreach ($a as $c) {  // Iterate over the array
  echo $c . " ";      // Print out (in order) 3 4 5
}
$o=new C();           // Create a new object
$o->m($i);            // Call method m
$s = <<<EOD           // Start multiline string
  A string
  with multiple lines
EOD;                  // End multiline string
\end{lstlisting}

\section{\ourvm}

\ourvm is a language composition of PHP and Python, implemented by composing
two existing RPython interpreters: \hippy (for PHP) and \pypy (for Python). \pypy is
an industrial-strength Python interpreter which can be used as a drop-in
replacement for CPython 2.7.8; \hippy is a partially complete PHP 5.4 interpreter.
\ourvm is a `semantics preserving' composition in the sense that it adds behaviour to
both Python and PHP, but does not alter or remove existing behaviour.

\ourvm programs currently start by executing PHP code. There
is no deep reason for this choice, and one could easily allow it to start by executing
Python code instead. Since they start as normal PHP programs, `raw' \ourvm programs
require \texttt{<?php~\ldots~?>} tags around the entire file. In the interests of brevity,
we omit these in all code listings.

We stage our explanation of
\ourvm as follows: the design and implementation of its FFI (Section~\ref{sec:ffi design});
its support for syntactic composition (Section~\ref{sec:syntactic}); and finally
cross-language variable scoping (Section~\ref{sec:xscope}).

\section{\ourvm FFI}
\label{sec:ffi design}

\ourvm defines an FFI which is the core upon which advanced functionality is
later built. A simple example of PHP code using the FFI to interact with Python is as
follows:
\begin{lstlisting}[label={lst:import_py_mod}, numbers=left]
$random = import_py_mod("random");<!\label{line:import_py_mod}!>
$num = $random->randrange(10, 20);
echo $num . "\n";<!\label{line:php_call_mod_members}!>
\end{lstlisting}
The code first imports Python's \texttt{random} module into PHP (line 1) before
calling the Python \texttt{randrange(x, y)} function to obtain a random integer between
10 and 20 (line 2) which is then printed (line 3). The
only explicit use of the FFI in this example is the call to \texttt{import\_py\_mod}. However, the FFI is implicitly
used elsewhere: the PHP integers passed as arguments to \texttt{randrange}
are converted to Python integers, and the result of the function is converted
from a Python integer to a PHP integer. As this may suggest, the FFI is two-way
and Python code can also call PHP.

\subsection{FFI Design}

Many parts of \ourvm's FFI are fairly traditional, while some are unusual due to
the semantic friction between Python and PHP. To the best of our knowledge there
has not previously been an FFI between Python and PHP, so our solutions are
necessarily novel.

\subsubsection{Data Type Conversions}
\label{sec:designconversions}

All FFIs have to define data type conversions between their constituents.
Since primitive data types in both PHP and Python are immutable, \ourvm
directly maps them from one language to the other (e.g.~a PHP integer is transformed
into a Python integer). Arbitrary user objects cannot be directly mapped and are
instead wrapped in
an \emph{adapter} which allows the other interpreter to work transparently
with the underlying foreign instance. A PHP object, for
example, appears to Python code as a normal Python object, whose attributes
and methods can be accessed, introspected etc. Passing an
adapter back to the language from which it was created simply removes
the adapter. Adapters are immutable, only ever pointing to single
object in their lifetime; the trace optimiser is then extremely
effective at removing the overhead of repeated adaptations.

Collection data types are more involved. Python separates the notion of a list
(i.e.~resizeable array) from that of a dictionary (i.e.~hashmap).
In contrast, PHP has a single dictionary type called, somewhat confusingly, an
array. PHP arrays are therefore also used wherever `lists' are required. This
presents an interesting case of semantic friction. Python lists
and dictionaries can both be sensibly adapted in PHP as arrays. PHP arrays
passed to Python, in contrast, are ambiguous: should they be adapted as
lists or dictionaries? It is easy to design schemes which can be dangerously
subverted. For example, a PHP array which `looks like' a list might seem best
adapted as a Python list, but later mutation to its keys (e.g.~adding a
string key) can turn it into something which is clearly not a list.

The only consistent
design is therefore to default to adapting PHP arrays as Python dictionaries.
However, users often know that a given PHP array is, and always will be,
equivalent to a list. Therefore, PHP arrays adapted as Python dictionaries
have an additional method \texttt{as\_list}, which re-adapts
the underlying array as a Python list. Whenever operations on
the list adapter are called, a check is made to see whether the underlying PHP array is
still list-like; if it is not (e.g.~because a non-integer key has been added) an
exception is raised.

In general, converting an adapted object to its `host' language simply requires
removing the adapter. The one exception is a Python list which has been passed
to PHP, adapted as a PHP array, and which is subsequently returned back to Python.
Since our data-conversion rules dictate that
PHP arrays are adapted as Python dictionaries, Python code expecting a PHP
array would get a surprising result if the PHP array returned was unwrapped
directly to a Python list (rather than a dictionary). Thus a Python list adapted as a PHP array and then
returned to Python has a special Python dictionary adapter. Only if \texttt{as\_list}
is called on that adapter is the underlying Python list returned.

\subsubsection{Mutability}

PHP data types are immutable except for objects (which are mutable in
the same way as objects in Python) and references. Immutable data types often
use copy-on-write semantics. For example, appending to an array
creates a copy with an additional element at the end.
Operations on references -- mutable cells which typically point to immutable
data -- are passed onto the underlying datum,
which may be replaced. For example, appending to a reference which
points to an array mutates the reference to point to the newly copied array.

\label{sec:arrayrefs}
Since it is common to wrap PHP arrays in a PHP reference, and since Python's
expectations are that such data types are mutable, \ourvm does not directly
adapt arrays: instead, arrays are replaced by references to arrays, which
are then adapted. Put another way, PHP arrays
always appear to Python as mutable collections, whether adapted as lists or
dictionaries, and those mutations are visible to PHP code as well.

\subsubsection{Cross-language Calls}
\label{sec:args}

Both Python and PHP functions can be adapted and passed to the other language where they can be
called naturally. There are, however, two cases of semantic friction: Python functions with keyword
arguments; and PHP's pass-by-reference mechanism.

Simplifying slightly, Python functions accept zero or more mandatory,
ordered arguments, and zero or more unordered,
keyword arguments, each of which has a default value. Function
calls must pass parameters for each ordered argument and then
zero or more keyword arguments. The following example shows such a
function and an example call:
\begin{lstlisting}[numbers=left]
def fmturi(host, path, scheme="http", frag="", query=""):
  uri = "%s://%s%s" % (scheme, host, path)
  if query: uri += "?%s" % query
  if frag: uri += "#%s" % frag
  return uri
fmturi("google.com", "/", frag="q=ecoop")
\end{lstlisting}
While PHP allows parameters to have default values, arguments
must be passed in order, and there is no notion of keyword arguments. To
enable PHP to call Python functions such as \texttt{fmturi},
\ourvm adds a global PHP function \texttt{call\_py\_func(f, a,
k)} where \texttt{f} is an adapted Python function, \texttt{a} is an array of regular arguments, and \texttt{k} is an
array of keyword arguments. From PHP, one can thus emulate the
function call from line 6 as follows:
\begin{lstlisting}[numbers=left]
call_py_func($fmturi, array("google.com", "/"), array("frag" => "q=ecoop")).
\end{lstlisting}
By default, PHP parameters are pass-by-value but function signatures can mark their
arguments as being pass-by-reference by prepending the parameter name with
\texttt{\&}. When a function with such a parameter is called, a reference
is created which points to the argument passed (if it is not already
a reference). Thus one can write a PHP function which swaps the
contents of the variables passed to it:\label{php swap}

\begin{lstlisting}[numbers=left]
function php_swap(&$x, &$y) {
  $tmp = $y;
  $y = $x;
  $x = $tmp;
}
$a = 10; $b = 20;
php_swap($a, $b);
echo "$a $b\n"; // prints "20 10"
\end{lstlisting}

\noindent As this example shows, the code calling \texttt{php\_swap}
has no control over whether it is passing arguments with pass-by-value or
pass-by-reference semantics---indeed, calling \texttt{php\_swap} updates
\texttt{\$a} and \texttt{\$b} so that they point to
the newly created references. Since \ourvm needs
to allow Python functions to be used as drop-in replacements for PHP functions, we need
a notion of pass-by-reference for Python function
arguments. This is tricky since Python has no explicit notion of a reference.

\label{php_decor}\label{phpref}
\ourvm takes a two-stage approach to reducing this case of semantic friction.
First, we introduce an explicit \texttt{PHPRef} adapter into
Python which represents a mutable PHP reference. \texttt{PHPRef}s support two explicit operations: \texttt{deref()}
returns the value inside the reference; and \texttt{store(x)} mutates the
reference to point to \texttt{x}. Second, we add a Python decorator
\texttt{php\_decor} which takes a keyword argument \texttt{refs} which specifies
(as a sequence of argument indices)
which arguments are pass-by-reference. With this, we can write a Python swap
function as follows:
\begin{lstlisting}[numbers=left]
@php_decor(refs=(0, 1))
def py_swap(a, b):
  tmp = a.deref()
  a.store(b.deref())
  b.store(tmp)
\end{lstlisting}
Although it may be tempting to think that \texttt{PHPRef}s should be transparent
in Python, as they are in PHP, we found such a scheme to be impractical: it
would require changing every possible part of the Python language and
implementation where one can read or write to variables. Explaining the
effects to users would be extremely challenging, as would changing the
implementation. In \pypy, for example, we estimate that this would involve changing
around 100 separate locations.

Because \texttt{PHPRef}s are explicit, calling a PHP function with
pass-by-reference arguments from Python is possible but, inevitably,
somewhat clunky. Pass-by-reference arguments must be explicitly passed a
\texttt{PHPRef} object; other object types lead to a run-time exception.
Thus Python can call \texttt{php\_swap} as follows:
\begin{lstlisting}[numbers=left]
xref, yref = PHPRef(x), PHPRef(y)
php_swap(xref, yref)
x, y = xref.deref(), yref.deref()
\end{lstlisting}

\subsection{\ourvm FFI Internals}
\label{sec:pyhyp_internals}

Until now, we have detailed the language \ourvm presents to the user.
We now consider \ourvm's internal implementation details.
\ourvm required modifying both \hippy and \pypy. We added
modules to both \hippy (\texttt{pypy\_bridge}) and \pypy (\texttt{hippy\_bridge}),
which encapsulate most of \ourvm's behaviour. Most of
the common behaviour resides in the \texttt{pypy\_bridge}  module, though it could just
as easily reside in \texttt{hippy\_bridge}. Some behaviour is implemented
by invasively modifying existing \hippy or \pypy code.

\subsubsection{Data Type Conversion}
\label{sec:conversions}

Viewed from a suitable level of abstraction, both \hippy and \pypy implement
their respective languages using broadly similar data type hierarchies: a root
data type class -- not entirely coincidentally, called \texttt{W\_Root} in both
interpreters -- from which all objects inherit.
Generally \ourvm adapters extend, directly or via a subclass,
one of the \texttt{W\_Root}s.

We added methods to both root data type classes: a \texttt{to\_py}
method to every PHP data type; and a \texttt{to\_php} method to every
Python data type. Calling \texttt{to\_py} on a PHP datum creates a Python adapter
(and vice versa for \texttt{to\_php}). The default implementations of
\texttt{to\_py} and \texttt{to\_php} return generic adapters, but other
data types override them to return specialised adapters. Calling \texttt{to\_py}
/ \texttt{to\_php} on an adapter simply returns the adapted datum. The only
exception is calling \texttt{to\_py} on a Python list which has been adapted as
a PHP array; rather than returning the Python list itself, \ourvm is forced to
return a special (new) variant of a Python dictionary (see
Section~\ref{sec:designconversions}).

The generic adapters -- one each in \hippy and \pypy -- simply forward attribute
lookups, method calls, and the
like onto the adapted object. \ourvm then defines a number of specialised
adapters: 10 additional Python adapters and 8 additional PHP adapters. Some of
the special adapters expose different behaviour to the user (e.g.~collection
data types), whereas some deal with low-level differences between
data types in the VM (e.g.~\pypy's layout requires separate adapters to be defined
for functions and methods). As \pypy uses storage strategies to optimise collection
data types~\cite{bolz13strategies}, adapted PHP collections
create Python collection instances that use \ourvm-specific strategies rather than
subclassing \texttt{W\_Root}.

The code for adapters is self-contained and relatively simple. Together, the PHP and
Python adapters are just under 1400LoC, of which 400LoC
implements new storage strategies.

\subsubsection{Mutability}

In order to make PHP arrays mutable from Python, \ourvm requires
PHP arrays passed to Python to be wrapped in a reference. However,
simply adding a reference when an adapter is created would lead to mutations from
Python not being seen by PHP. \ourvm therefore handles arrays specially:
the following two examples give a flavour of this.

The \texttt{ARG\_BY\_PTR} PHP opcode organises function arguments prior to a
function call, and is where arrays passed to a call-by-reference function
argument have their storage in the PHP frame replaced by a reference. \ourvm
adds a special case to this opcode: every PHP array passed to a Python function
is treated as if it was being passed to a call-by-reference function argument. This
ensures that the PHP frame observes mutations from Python.

Similarly, when Python loads an array from an adapted PHP object's attribute, \ourvm
must replace the attribute with a reference. This ensures that
the parent object observes mutations from Python. Implementing this is fairly
simple, as we reuse an existing function in \hippy
which can lookup an attribute and turn non-references into references (used to
implement PHP's standard \texttt{x=\&\$y->z} behaviour). We add a new flag to
this function, since we only want to turn arrays (and not other data types) into
references.

\subsubsection{Cross-language Calls}

For the most part, cross-language calls are simple to implement. \ourvm is
careful to ensure that the necessary glue code is optimised and does not
obstruct meta-tracing's natural cross-language inlining. Most importantly,
this requires annotating the relevant RPython functions as being unrollable,
so that they dynamically specialise themselves to the number of
parameters passed by the user.

The \texttt{php\_decor} decorator is implemented as a normal Python (not
RPython) class. When the decorator is applied with the \texttt{refs} keyword
argument, the argument indices are stored in a normal, user visible attribute of the function
object. When the function is adapted for PHP, the adapter loads the
indices, which are later used by the PHP interpreter to determine which parameters are
to be passed by reference.

\subsubsection{Transparency}

Ensuring that adapters are as transparent as possible inevitably requires invasive modifications of
\hippy and \pypy. We were helped by the fact that both interpreters
centralise all the behaviour we wished to modify. For example,
implementing identity transparency in \pypy is easy, as identity
checks are not handled directly but handed over to an object's
\texttt{is\_w} method which compares the current object with another
for identity equality. \ourvm adapters override this method so that if
the objects being compared are both adapters of the same type, the identity check is forwarded on to the
underlying adapted objects. The \texttt{is\_w} method of \texttt{W\_PHPGenericAdapter}
(the class representing an adapted PHP object in \pypy)
shows this idiom:
\begin{lstlisting}[numbers=left]
def is_w(self, other):
  if isinstance(other, W_PHPGenericAdapter):
    return self.w_php_obj is other.w_php_obj
  return False
\end{lstlisting}

\section{Syntactic Composition}
\label{sec:syntactic}

Traditionally, FFIs have made the implicit assumption that the source
code of each language involved is kept in different files. In this section, we
show how \ourvm allows PHP and Python code to be used within a single file. To avoid tedious
duplication of explanation, we concentrate our explanation on embedding
Python into PHP, though \ourvm also allows PHP to be embedded into Python.

\subsection{Functions}
\label{the embed functions}

At a low-level, \ourvm provides simple support for embedding Python inside
PHP (and vice versa) as strings. For example, the \texttt{compile\_py\_func} function takes a string
containing a single Python function and returns a Python function object
that is adapted as a callable PHP instance. The following example
embeds a Python function inside PHP and calls it to produce
a random number between 0 and 10:
\begin{lstlisting}[numbers=left]
$src = <<<EOD
def randnum(n):
  import random
  return random.randrange(n)
EOD;
$randnum = compile_py_func($src);
echo $randnum(10) . "\n";
\end{lstlisting}
In this paper we mostly use \texttt{compile\_py\_func},
though \texttt{compile\_py\_func\-\_global} can be used to compile
a Python function and put it in PHP's global function namespace.

Although the \texttt{compile\_*} functions are called at run-time, they are
surprisingly fast. First, due to PHP and Python being simple languages to
compile, both \hippy and \pypy have efficient compiler implementations.
Second, when JIT compiled, \ourvm caches the bytecode output of \texttt{compile\_*}
calls. Re-evaluating a function thus produces a new (cheap) function object
while reusing the (expensive) bytecode object which underlies it.

\subsection{Methods}
\label{sec:methodemb}

\ourvm supports inserting Python methods into PHP classes with the
\texttt{compile\_py\_meth(c, f)} function, where the Python method source \texttt{f} is
compiled and inserted into the class named by the string \texttt{c}. However,
this feature must be used carefully because \hippy's implementation takes advantage of
the fact that PHP's classes can be statically compiled. This means that,
by the time a program has started running, normal PHP classes cannot
be altered. This poses a problem for \ourvm because its syntactic
embedding is currently performed at run-time. We work around this by enclosing a
PHP class inside curly braces, which delays its
compilation until run-time. We also require that all
calls to \texttt{compile\_py\_meth(c, ...)} immediately follow the
definition of \texttt{c} and that the class and all such calls be surrounded by
curly brackets. This also affects sub-classes of \texttt{c}, whose execution
must also be delayed by curly brackets (though not necessarily at the same
point in the source code).

PHP supports a Java-like public/private/protected method access scheme, but
Python has no concept of private or protected
methods.\footnote{Python attributes prefixed by two underscores have their
names mangled, but are otherwise publicly accessible.} To model this,
\ourvm extends the \texttt{php\_decor} decorator (see
Section~\ref{php_decor}) with an optional extra
argument \texttt{access} which accepts a string access value (\texttt{"public"},
\texttt{"private"}, or \texttt{"protected"}). Similarly, the
option \texttt{static} boolean argument allows static methods to be denoted.
These arguments can be combined to specify
that a Python method is e.g.~protected and static.

\subsection{Using Eco to Express Embeddings}
\label{sec:eco}

\begin{figure}
\begin{lstlisting}[caption={Composing PHP and Python grammars in Eco.
    \texttt{Grammar(n, p)} loads a grammar named \texttt{n} from path \texttt{p}.
    \texttt{change\_start} changes the start rule of a grammar.
    \texttt{g1.add\_alternative(r, g2)} adds a new alternative to the rule
    \texttt{r} in grammar \texttt{g1} to the start rule of grammar
\texttt{g2}.}, label={lst:grammars}, numbers=left]
# Create PHP grammar referencing Python (and vice versa)
python = Grammar("python.eco")
php = Grammar("PHP+Python", "php.eco")
python.add_alternative("atom", php)
php.add_alternative("top_statement", python)
php.add_alternative("class_statement", python)
php.add_alternative("expr", python)
# Create Python expressions-only grammar
python_expr = Grammar("Python expressions", "python.eco")
python_expr.change_start("simple_stmt")
php.add_alternative("expr", python_expr)
\end{lstlisting}
\end{figure}

For simple uses, the \texttt{compile} functions are tolerable as-is, but
they tend to obfuscate embedded code, especially in multi-level embeddings
(e.g.~PHP inside Python inside PHP), where string escaping becomes onerous.

In order to make \ourvm's syntactic composition more palatable, we make
use of the Eco editor~\cite{diekmann14eco}. Eco allows users to compose grammars
and to write composed programs. In essence, one takes a
context-free grammar $X$ and adds a reference from a rule $R \in X$
to another grammar $Y$. When entering
language $X$ into Eco, the user can switch to entering language $Y$ by creating a
language box. Language boxes delineate when one
language ends and another starts, but do not introduce ambiguity or complexity into a parser.
Eco parses as the user type, and thus always has access to
parse trees for all languages involved in a composition. However, from
the user's perspective, editing in Eco feels like a normal text
editor, except when creating a language box or moving between nested language boxes (a
fairly rare occurrence).

We first had to enable users to write \ourvm programs. Since Eco comes with a Python
grammar, we only had to add a PHP grammar (which we adapted, with
minor modifications, from Zend PHP) and a PHP lexer. We then used Eco's
Python interface to express the join points between the two
grammars (see Listing~\ref{lst:grammars}). Simplifying slightly, in PHP one
can add Python language boxes wherever PHP statements or expressions are
valid; and in Python one can add PHP language boxes wherever a Python
expression is valid. From the users perspective, this means that when using Eco for \ourvm,
they create a new \texttt{PHP+Python} file and start typing PHP. When they
want to insert a Python function they insert a \texttt{Python+PHP} language box;
when they want to insert a Python expression they insert a \texttt{Python
expression} language box.

We then had to implement an exporter from parse trees relative to the composed \texttt{PHP+Python}
composed grammar to \ourvm-compatible input. The exporter automatically inserts
\texttt{compile\_*} functions, follows \texttt{compile\_py\_meth}'s
restrictions, and escapes arbitrarily deeply nested language boxes.  For
example, the \texttt{gen} Python language box in the \texttt{RNG} class in
Figure~\ref{fig:rngexample} is exported as follows:
\begin{lstlisting}[numbers=left]
{
class RNG { ... }
compile_py_meth("RNG", "def gen(self, amount):\n \\
                        while amount > 0:\n \\
                          amount -= 1\n \\
                          yield self.pcg8_random()");
}
\end{lstlisting}
Figure~\ref{fig:rngexample}'s second language
box (line 20) is more interesting. \ourvm has no explicit interface for
embedding Python expressions. Instead, the expression is encoded as a callable PHP
instance which is compiled and immediately called:
\begin{lstlisting}[numbers=left]
$l = call_py_func(compile_py_func("f = lambda: [x % 64 for x in rng.gen(25)];"));
\end{lstlisting}
In the rest of the paper we use the term `language box' to refer to an embedded
language fragment irrespective of whether Eco is used or not.

\subsubsection{Exceptions}

When a native exception crosses the language boundary, \ourvm adapts it, before
re-raising the exception. For example, if PHP calls Python code which raises the
Python exception \texttt{ZeroDivisionError}, then the exception will appear to
PHP as a generic \texttt{PyException}. As with all other adapters, an adapted
exception crossing back to
its native language (e.g.~a \texttt{PyException} which percolates back to
Python) simply has its adapter removed.

However, adapters on their own do not solve a crucial problem: cross-language
stacktraces. By default, both \hippy and \pypy can only print out their own
frames in stacktraces, which makes cross-language exceptions seem to appear out
of thin air. Equally frustratingly, frames which represent inner language boxes
report incorrect line numbers, as each language box's frame assumes it
starts at line 1.

\begin{figure}[t]
\centering
\includegraphics[width=0.95\textwidth]{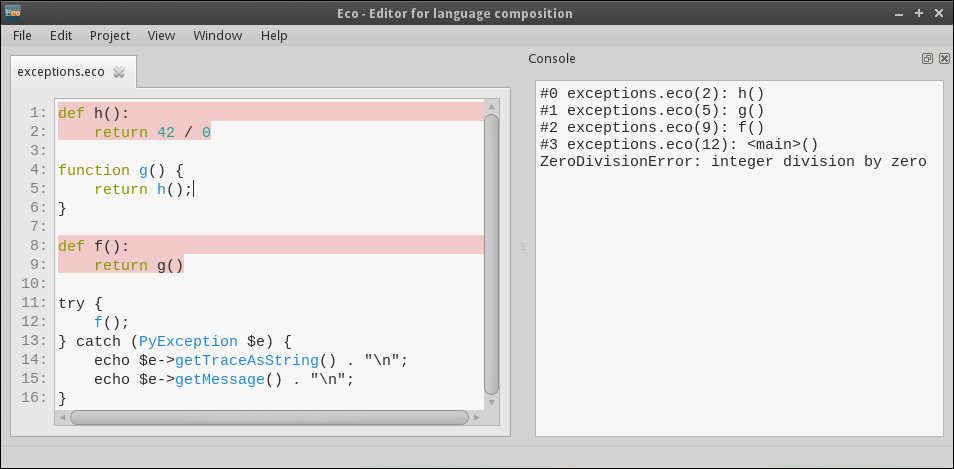}
\caption{Cross-language exception handling in \ourvm, showing
that the stacktrace presents entries from within language boxes correctly. As an example,
the first line of the stacktrace should be read as follows: ``The 0th
entry in the stacktrace relates to the exceptions.eco file, line 2,
within the \texttt{h} function''}
\label{fig:exns}
\end{figure}

\ourvm fixes both problems. First, we altered
\hippy and \pypy's stacktrace functions to call each other for their
appropriate frames. Second, we added two arguments to \ourvm's compilation functions
(e.g.~\texttt{compile\_py\_func}) to allow Eco to pass the file and line
offset the compiled text is relative to. The stacktrace routines then use
this information to adjust the reported locations. The end result is that
cross-language stacktraces are just as informative as mono-language
stacktraces, as can be seen in Figure~\ref{fig:exns}.

\section{Cross-language Scoping}
\label{sec:xscope}

\begin{figure}[t]
\centering
\includegraphics[width=.46\textwidth]{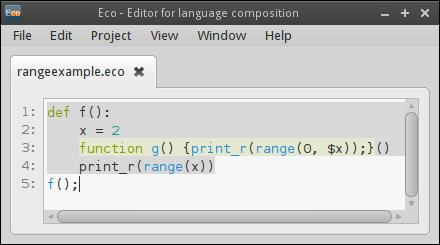}
\caption{Cross-language scoping with nested language boxes: each language
expects to see its own
global scope. Lexical scoping suggests that \texttt{x}
referenced on lines 3 and 4 should bind to the definition on line 2. Since
PHP and Python's \texttt{range} functions are incompatible,
the PHP language box on line 3 must reference a different
\texttt{range} function to that on line 4. However,
both the PHP and Python language boxes wish to use PHP's \texttt{print\_r}
function on lines 3 and 4. \ourvm's
scoping rules respect these desires and this example prints out 0, 1, 2, 0, 1.}
\label{fig:range}
\end{figure}

For syntactic composition to be useful, we believe that users must be able to
reference variables across language boxes. \ourvm therefore allows both Python
and PHP to reference variables in the other language, making syntactic
composition significantly more powerful and usable. This raises a novel design
challenge: what are sensible cross-language scoping rules? The major
challenge is to deal with both language's expectations surrounding global
namespaces. We eventually settled upon scoping rules that are relatively easily explained, and which
impose a mostly lexical system. This is no small matter, as PHP
and Python have significantly different scoping rules.

In this section, we first define a simplified version of the scoping rules in
each language (since PHP and Python share neither common
semantics nor terminology, we do our best to homogenise the description).
We then define \ourvm's additional rules for PHP referencing Python
and vice versa, before explaining how these rules are implemented.

\subsection{PHP and Python's Namespace Semantics}
\label{scope:php namespace semantics}

PHP has separate global namespaces for functions, classes, constants, and variables.
Any given name can appear in multiple namespaces, as syntactic
context uniquely identifies which namespace is being used
(e.g.~\texttt{new x()} references a class, whereas
\texttt{x()} references a function). PHP has no concept of modules,
textually `including' files in similar fashion to C headers. Thus the
global namespaces span all PHP files.\footnote{Although PHP 5.3 introduced a
mechanism for defining non-global namespaces, this does not effect our explanation.}
The namespaces for functions, classes, and constants can have new names added to
them dynamically, but existing names cannot be removed or changed. In contrast,
names can be added or removed to the namespace for variables
at will. Each function then defines a local scope; variable lookups within a
function first search the local scope before searching the global variable
namespace.

Python modules each have a single `global' namespace, to which names can be
added and removed. Functions define their own local namespace. Lexical
lookups (for names defined in the current or parent function's scope)
are determined statically (the set of local names cannot be
modified); global lookups are performed dynamically.

Neither PHP nor Python has Scheme-esque closures\footnote{For our purposes, we
consider what PHP calls a `closure' to be a first-class anonymous function.}:
nested functions can read a parent function's variables but writes
are not shared between the two. In the interests of brevity, we consider
PHP and Python's variable scoping rules to be local
and global, but not lexical.

\subsection{\ourvm's Cross-language Scoping Rules}

The use of cross-language scoping in Figure~\ref{fig:rngexample} may suggest
that cross-language scoping design is a matter of applying traditional language
design principles. Figure~\ref{fig:range} shows a more challenging example,
where modern expectations about lexical scoping and PHP and Python's global
namespaces appear to clash.

\ourvm resolves this issue with the following simple design. First, PHP
and Python's local scoping rules remain unchanged. Second, we split the
search for variables which are not bound in
the current language box into two distinct phases: the `recursive' phase,
and the `global' phase. The recursive phase
searches language boxes (from inner to outer) for a matching
variable definition. Failing this, the global phase searches the
global namespace(s) of the current language box's language; if no match
is found, it then searches the global namespace(s) of the other language.

The recursive phases for PHP and Python are conceptually identical. Python's
global phase is simple, but PHP's is
complicated by its multiple global namespaces. If the search originated from PHP, then
only the appropriate namespace for the syntactic context is used (e.g.~if,
syntactically, a function was looked up, only the function namespace is
searched); if the search originated in Python the namespaces are searched in the
following order: global functions, classes, then constants.

Performance reasons led us to make one small adjustment to the PHP global
phase search. PHP has the ability to lazily load classes; every failed class lookup
triggers a (fairly slow) check for user-defined lazy loading mechanisms. In mono-PHP
this is a sensible mechanism, as in practice either a class is found in the
namespace, or it is lazily loaded, or an error is raised. However, since
cross-language scoping frequently checks for the existence of names that
do not, and will never, exist, the cumulative performance effect
can be frustrating. Since disabling lazy loading would break many existing
PHP applications, we tweaked \ourvm's scoping rules. The search for a Python name in PHP's
global namespaces is `sticky': if a name $x$ was found to be (say) a class on
the first search in a given scope, it will only ever return a class on
subsequent searches (i.e.~if a function $x$ is later added to the functions
namespaces, it will not be returned as a match in that scope). This small
loss of dynamicity increases performance in some benchmarks by around 50\%.

Using Figure~\ref{fig:range} as a concrete example, we can see how these rules
apply in practice. First, consider the PHP variable reference \texttt{\$x} on
line 3. There is no binding of \texttt{x} in the PHP language box so when the
code is executed a recursive phase search commences: the
parent (Python) language box is inspected and a binding found on line 2.
The \texttt{range} function reference on
line 3 starts with the same pattern: a recursive phase search looks in the Python
language box for a binding but fails. A PHP global phase search then commences; since \texttt{range} was
syntactically referenced as a function, only the global function namespace is
searched, where a match is then found. The
\texttt{print\_r} function reference on line 3 follows the same pattern. The reference to a
\texttt{range} function in the Python language box at line 4 starts with a recursive
phase search which looks into the parent PHP language box's for a suitable
binding (i.e.~a name starting with a `\texttt{\$}') but fails. It then does a global
phase search in Python's global namespace, finding Python's built-in \texttt{range}
function. The \texttt{print\_r} reference on line 4 is more interesting. A
recursive phase search fails to find a match. A global phase search then searches in
Python's global namespace and fails before trying PHP's global namespaces,
starting with functions, where a match is found.

\ourvm's scoping rules also work well in corner-cases (e.g.~\ourvm deals
with PHP's superglobals sensibly). Note that the scoping rules of both languages
are partly or wholly
dynamic: that is, in some situations, bindings can be changed at run-time.
\ourvm's scoping rules maintain PHP's and Python's dynamic lookup properties
since some programming idioms (particularly in PHP, but also
in Python) rely on adding or removing bindings.

\subsubsection{Implementation}
\label{sec:implementing scopes}

To add \ourvm's scoping rules to \hippy and \pypy, we first needed to
connect language box scopes together at run-time, and then intercept the
locations where PHP and Python global variables are read and written to.

Connecting language box scopes is made relatively simple by the fact that each
is constructed by a \texttt{compile\_*} function (see Section~\ref{the embed
functions}). The outer part of any \ourvm program is, by definition, a PHP
language box and every other language box is nested inside that. Thus
any call to \texttt{compile\_py\_*} implicitly receives a reference
to the PHP frame from which it was called. The reference is
then stored in a \texttt{PHPScope} object,
which \ourvm attaches to the Python function object being
created; nested Python functions inherit a \texttt{PHPScope} object
from their parent function, so that multiply nested functions can still access
outer language boxes. Similarly, when a PHP language box is nested inside Python,
a \texttt{PyScope} object is created and placed inside a PHP function's
bytecode object. This simple scheme means that, from any PHP or
Python frame, one can walk a chain from the current to the outermost language box.

To actually search outer language box's scopes, we have to modify those parts
of \hippy and \pypy which perform global lookups. In \hippy, we modify the
3 separate functions on the main interpreter object which perform searches of
functions, classes, and constants, as well as the \texttt{lookup\_deref}
function on frames which lookups up variables. An elided version of the
\texttt{locate\_function} function -- which searches for a PHP function
\texttt{n} -- shows the small scale of such modifications:
\begin{lstlisting}[numbers=left]
def locate_function(n):
  py_scope = self.topframeref().bytecode.py_scope
  if py_scope is not None:
    ph_v = py_scope.ph_lookup_local_recurse(name)
    if ph_v is not None: return ph_v
    ph_v = self.lookup_function(name)
    if ph_v is not None: return ph_v
    ph_v = py_scope.ph_lookup_global(name)
    if ph_v is not None: return ph_v
  else:
    func = self.lookup_function(name)
    if func is not None: return func
  self.fatal("Call to undefined function %s()" % name)
\end{lstlisting}
In essence, the original function consisted of lines 11 and 12; \ourvm adds
lines 1--9. When a PHP function performs a global function lookup, and
that PHP function is nested inside a Python language box (lines 1 and 2)
then a local phase search is performed (lines 4 and 5). If unsuccessful, the global
phase search then commences: first the PHP global function namespace is searched
(lines 6 and 7) before Python's global namespace is searched (lines 8 and 9).
The modifications made to \pypy are identical in idiom, modifying
the two opcodes (\texttt{LOAD\_GLOBAL} and
\texttt{STORE\_GLOBAL}) which read and write global variables.

Since \ourvm's rules are highly dynamic, we rely heavily on use of Self-style
maps (see~\cite{chambers89efficient}) to turn most name lookups from (slow)
dictionary lookups of names into (fast) constant lookups.
\hippy originally made only limited use of maps in its global
namespaces: we altered it to use maps extensively.\footnote{This optimisation
also helps plain \hippy, as it significantly improves the performance of
programs such as MediaWiki and phpBB that use the \texttt{\$GLOBALS} superglobal.}
We also used maps for the sticky name\-space search. Looking up a
variable in a global phase search in a PHP scope returns an integer representing
unknown (i.e.~this name has not previously been searched for in this context), class, function,
constant, or not found (i.e.~a search was previously done and no match was
found). After tracing, virtually all global lookups turn
into a small number of (quick) attribute guards, avoiding (slow)
dictionary lookups.

\section{Experiment}
\label{sec:method}

To understand \ourvm's performance characteristics, we define three classes
of benchmarks: small microbenchmarks, large microbenchmarks, and permutation benchmarks.
Benchmarks come in four variants: mono-language PHP (henceforth `mono-PHP');
mono-language Python (`mono-Python'); composed PHP and Python where PHP is the
`outer' language (`composed-PHP') and where Python is the
`outer' language (`composed-Python'). We run these benchmarks
on several PHP and Python implementations.

\subsection{Benchmarks}

\label{microbenchmarks}

Our small microbenchmarks, focus on single language features in isolation, and are
useful for identifying low-level pinch points. Each of our small microbenchmarks
consist of two parts. In most, an outer loop repeatedly
calls an inner function (e.g.~the \emph{total\_list} benchmark's inner
function takes a list of integers and sums them). In the remainder,
an outer function generates elements, and an inner function
consumes them. In the composed variants, the inner
and outer components are implemented in different languages.

Some small microbenchmarks cannot be implemented in all variants. The
\emph{ref\_swap} benchmark measures the performance of operations on PHP
references and \texttt{PHPRef}s (see Section~\ref{phpref}) and thus has no
mono-Python variant. Benchmarks which require putting PHP methods into
Python classes are currently not supported by \ourvm.
The complete list of small microbenchmarks can be found in
Appendix~\ref{sec:microbenchmarks}.

\begin{figure*}[t]
\begin{center}
\includegraphics[width=1.2\textwidth, center]{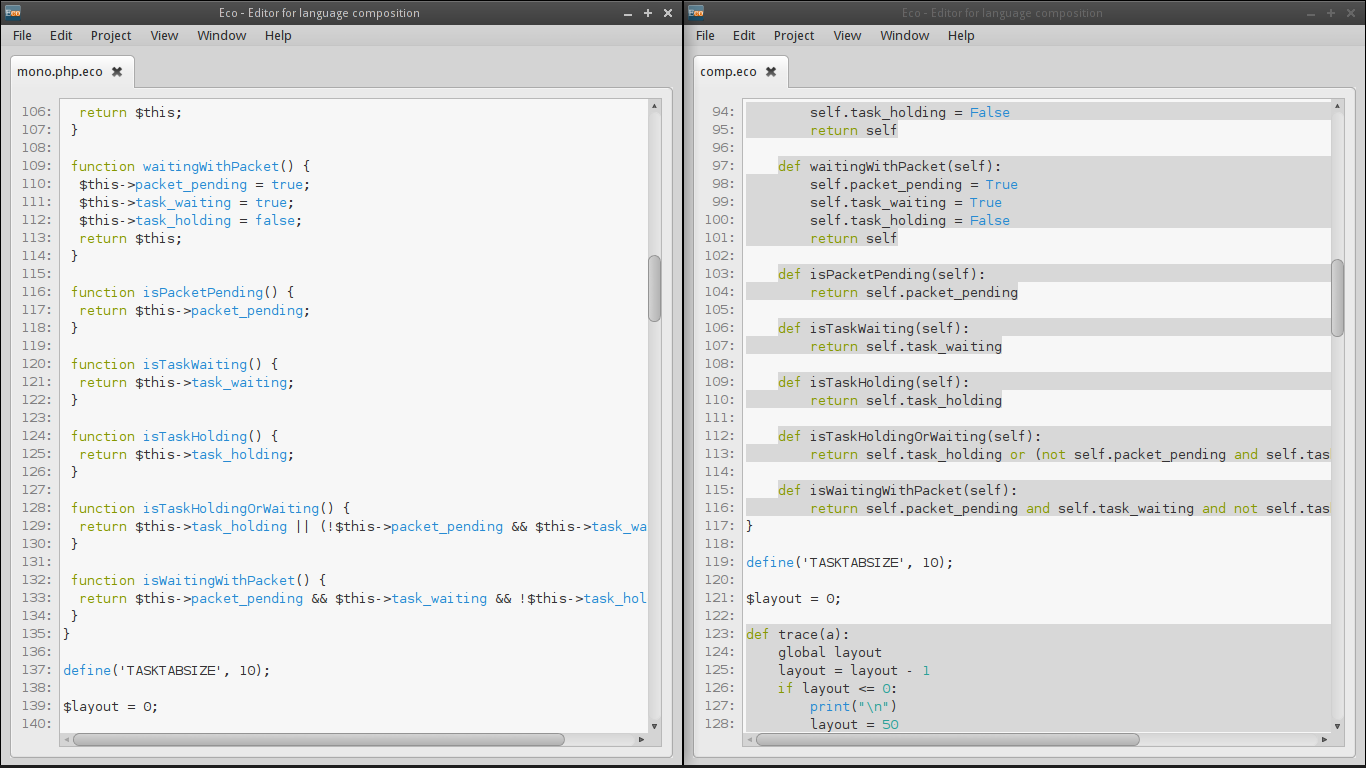}
\end{center}
\caption{The mono-PHP and composed-PHP variants of Richards side by side.
The composed-PHP variant of the benchmark contains a `shell' PHP
program with PHP classes whose methods are Python language boxes.
Global variables remain defined in PHP, so that the benchmarks also include
an element of cross-language scoping.}
\label{fig:richards}
\end{figure*}

Our large microbenchmarks aim to measure performance more broadly.
We use four `classic' benchmarks: \emph{Fannkuch} counts permutations by
continually flipping elements in a list~\cite{anderson94fannkuch}; \emph{Mandel}
plots an ASCII representation of the Mandelbrot set into a string
buffer\footnote{\texttt{Zend/bench.php} in the Zend distribution of PHP.};
\emph{Richards} models an operating system task
dispatcher\footnote{\url{http://www.cl.cam.ac.uk/~mr10/Bench.html}}; and
\emph{DeltaBlue} is a constraint solver~\cite{sannella93deltablue}.
To create composed variants of these benchmarks, we took the
mono-language variants and replaced each function with an
implementation in the other language.
The composed-PHP variants of Richards and DeltaBlue are thus PHP
`shell' classes containing many Python methods (33 and 75
respectively), with variables referenced between languages and data
repeatedly crossing from PHP to Python (Figure~\ref{fig:richards}
shows the mono and composed-PHP variants of Richards
alongside each other). In other words, Richards and DeltaBlue
are designed to heavily test \ourvm's cross-language performance. In contrast,
Fannkuch is a single function, and so the composed variant consists of a single
Python function embedded in PHP. This serves as a rough baseline, since we would
expect that the composed variant has roughly the same performance as \pypy.
Mandel also
started off life as a single function, but we made modifications to
make a more interesting benchmark: we split the innermost loop into a
separate function; made the function's parameters pass-by-reference; and
made the
function modify these references during execution. Since Mandel uses references,
there is no mono-Python variant. There is no composed-Python
variant of either Richards or DeltaBlue, since PHP methods cannot yet appear
inside Python classes.

The permutation benchmarks are designed to uncover whether some parts of a program are
faster in one or other language. Using the mono-PHP DeltaBlue benchmark as a
base, we created 79 permutations, each with one PHP function replaced by a
Python equivalent. We then
compare the timings of each permutation to the original mono-PHP benchmark. For brevity, we
henceforth refer to permutation number $x$ as $p_x$.

\subsection{Methodology}

\begin{table}[t]
\begin{center}
\begin{tabular}{ll}
\toprule
Interpreter&	Version(s)\\
\midrule
CPython	&2.7.10\\
HHVM		&3.4.0\\
\hippy		&git \#2ae35b80\\
\pypy		&2.6.0\\
Zend		&5.5.13\\
\ourvm		&Based on above \hippy/\pypy versions.\\
\bottomrule
\\
\end{tabular}
\caption{The VM versions used in this paper.}
\label{tab:versions}
\end{center}
\end{table}

Each benchmark was run on the following VMs (in alphabetical order): CPython, the standard
interpreter for Python; HHVM, a JIT compiling VM for PHP; \hippy; \ourvmcomp, which is \ourvm
running composed-PHP variants; \ourvmcompr, which is \ourvm running composed-Python
variants; \ourvmmono, which is \ourvm running mono-PHP
variants; \pypy; and Zend, the standard interpreter for PHP. The versions used
for each of these VMs is shown in Table~\ref{tab:versions}. Note that
\ourvmcomp, \ourvmcompr and \ourvmmono are all the save VM, and we use the
terms to be clear about which benchmarks \ourvm is running.

\label{no warmup chopping}
For each benchmark and VM pair, we ran 5 fresh processes, with each process
running 50 iterations of the benchmark. We then
used the bootstrapping technique described in \cite{kalibera12quantifying} to
derive means and 99\%{} confidence intervals for each pairing. Since we could not always
determine when a given VM had warmed up, we made no attempt to remove any
iterations from the process: thus our timings include warmup. All timings are
wall-clock with a sub-microsecond resolution.

All experiments were run on an otherwise idle 4GHz Core i7-4790
CPU and 32GiB RAM machine, running Debian 8. We disabled turbo mode and hyper-threading
in the BIOS. We used a tickless Linux kernel,
disabled Intel p-states, and ensured that the CPU
governor was set to maximum performance mode. All VMs were built with the system
GCC (4.9.2). We did not interfere with the garbage collection of any of the VMs,
which run as normal.

\subsection{Results}
\label{sec:results}
\begin{table*}[t]
\centering

\begin{adjustbox}{width=1.2\textwidth,center}
\begin{tabular}{lrrrrrrrr}
\toprule
Benchmark&\makebox[1.45cm][r]{CPython}&\makebox[1.45cm][r]{HHVM}&\makebox[1.45cm][r]{HippyVM}&\makebox[1.45cm][r]{PyHyp$_\textrm{PHP}$}&\makebox[1.45cm][r]{PyHyp$_\textrm{Py}$}&\makebox[1.45cm][r]{PyHyp$_\textrm{mono}$}&\makebox[1.45cm][r]{PyPy}&\makebox[1.45cm][r]{Zend}\\
\toprule
instchain&$\substack{\mathmakebox[.7cm][r]{33.451}\\{\mathmakebox[.7cm][r]{\scriptscriptstyle \pm 0.0679}}}$&$\substack{\mathmakebox[.7cm][r]{9.547}\\{\mathmakebox[.7cm][r]{\scriptscriptstyle \pm 0.0096}}}$&$\substack{\mathmakebox[.7cm][r]{0.912}\\{\mathmakebox[.7cm][r]{\scriptscriptstyle \pm 0.0011}}}$&$\substack{\mathmakebox[.7cm][r]{1.000}\\{\mathmakebox[.7cm][r]{\scriptscriptstyle }}}$&&$\substack{\mathmakebox[.7cm][r]{1.116}\\{\mathmakebox[.7cm][r]{\scriptscriptstyle \pm 0.0012}}}$&$\substack{\mathmakebox[.7cm][r]{0.675}\\{\mathmakebox[.7cm][r]{\scriptscriptstyle \pm 0.0007}}}$&$\substack{\mathmakebox[.7cm][r]{36.471}\\{\mathmakebox[.7cm][r]{\scriptscriptstyle \pm 0.1577}}}$\\
\addlinespace
l1a0r&$\substack{\mathmakebox[.7cm][r]{86.017}\\{\mathmakebox[.7cm][r]{\scriptscriptstyle \pm 0.0179}}}$&$\substack{\mathmakebox[.7cm][r]{4.052}\\{\mathmakebox[.7cm][r]{\scriptscriptstyle \pm 0.0020}}}$&$\substack{\mathmakebox[.7cm][r]{1.368}\\{\mathmakebox[.7cm][r]{\scriptscriptstyle \pm 0.0004}}}$&$\substack{\mathmakebox[.7cm][r]{1.000}\\{\mathmakebox[.7cm][r]{\scriptscriptstyle }}}$&$\substack{\mathmakebox[.7cm][r]{1.360}\\{\mathmakebox[.7cm][r]{\scriptscriptstyle \pm 0.0003}}}$&$\substack{\mathmakebox[.7cm][r]{1.359}\\{\mathmakebox[.7cm][r]{\scriptscriptstyle \pm 0.0003}}}$&$\substack{\mathmakebox[.7cm][r]{1.340}\\{\mathmakebox[.7cm][r]{\scriptscriptstyle \pm 0.0106}}}$&$\substack{\mathmakebox[.7cm][r]{38.778}\\{\mathmakebox[.7cm][r]{\scriptscriptstyle \pm 0.0078}}}$\\
\addlinespace
l1a1r&$\substack{\mathmakebox[.7cm][r]{83.803}\\{\mathmakebox[.7cm][r]{\scriptscriptstyle \pm 0.1407}}}$&$\substack{\mathmakebox[.7cm][r]{2.980}\\{\mathmakebox[.7cm][r]{\scriptscriptstyle \pm 0.0038}}}$&$\substack{\mathmakebox[.7cm][r]{1.306}\\{\mathmakebox[.7cm][r]{\scriptscriptstyle \pm 0.0017}}}$&$\substack{\mathmakebox[.7cm][r]{1.000}\\{\mathmakebox[.7cm][r]{\scriptscriptstyle }}}$&$\substack{\mathmakebox[.7cm][r]{1.303}\\{\mathmakebox[.7cm][r]{\scriptscriptstyle \pm 0.0016}}}$&$\substack{\mathmakebox[.7cm][r]{1.303}\\{\mathmakebox[.7cm][r]{\scriptscriptstyle \pm 0.0016}}}$&$\substack{\mathmakebox[.7cm][r]{1.140}\\{\mathmakebox[.7cm][r]{\scriptscriptstyle \pm 0.0022}}}$&$\substack{\mathmakebox[.7cm][r]{39.111}\\{\mathmakebox[.7cm][r]{\scriptscriptstyle \pm 0.1272}}}$\\
\addlinespace
lists&$\substack{\mathmakebox[.7cm][r]{8.047}\\{\mathmakebox[.7cm][r]{\scriptscriptstyle \pm 0.0139}}}$&$\substack{\mathmakebox[.7cm][r]{0.931}\\{\mathmakebox[.7cm][r]{\scriptscriptstyle \pm 0.0036}}}$&$\substack{\mathmakebox[.7cm][r]{0.975}\\{\mathmakebox[.7cm][r]{\scriptscriptstyle \pm 0.0020}}}$&$\substack{\mathmakebox[.7cm][r]{1.000}\\{\mathmakebox[.7cm][r]{\scriptscriptstyle }}}$&$\substack{\mathmakebox[.7cm][r]{0.560}\\{\mathmakebox[.7cm][r]{\scriptscriptstyle \pm 0.0012}}}$&$\substack{\mathmakebox[.7cm][r]{0.978}\\{\mathmakebox[.7cm][r]{\scriptscriptstyle \pm 0.0021}}}$&$\substack{\mathmakebox[.7cm][r]{0.497}\\{\mathmakebox[.7cm][r]{\scriptscriptstyle \pm 0.0010}}}$&$\substack{\mathmakebox[.7cm][r]{14.626}\\{\mathmakebox[.7cm][r]{\scriptscriptstyle \pm 0.0377}}}$\\
\addlinespace
ref\_swap&&$\substack{\mathmakebox[.7cm][r]{8.393}\\{\mathmakebox[.7cm][r]{\scriptscriptstyle \pm 0.0006}}}$&$\substack{\mathmakebox[.7cm][r]{1.000}\\{\mathmakebox[.7cm][r]{\scriptscriptstyle \pm 0.0002}}}$&$\substack{\mathmakebox[.7cm][r]{1.000}\\{\mathmakebox[.7cm][r]{\scriptscriptstyle }}}$&$\substack{\mathmakebox[.7cm][r]{0.700}\\{\mathmakebox[.7cm][r]{\scriptscriptstyle \pm 0.0001}}}$&$\substack{\mathmakebox[.7cm][r]{1.000}\\{\mathmakebox[.7cm][r]{\scriptscriptstyle \pm 0.0001}}}$&&$\substack{\mathmakebox[.7cm][r]{53.320}\\{\mathmakebox[.7cm][r]{\scriptscriptstyle \pm 0.0040}}}$\\
\addlinespace
return\_simple&$\substack{\mathmakebox[.7cm][r]{110.409}\\{\mathmakebox[.7cm][r]{\scriptscriptstyle \pm 0.1104}}}$&$\substack{\mathmakebox[.7cm][r]{7.049}\\{\mathmakebox[.7cm][r]{\scriptscriptstyle \pm 0.0019}}}$&$\substack{\mathmakebox[.7cm][r]{1.000}\\{\mathmakebox[.7cm][r]{\scriptscriptstyle \pm 0.0001}}}$&$\substack{\mathmakebox[.7cm][r]{1.000}\\{\mathmakebox[.7cm][r]{\scriptscriptstyle }}}$&$\substack{\mathmakebox[.7cm][r]{0.778}\\{\mathmakebox[.7cm][r]{\scriptscriptstyle \pm 0.0001}}}$&$\substack{\mathmakebox[.7cm][r]{1.000}\\{\mathmakebox[.7cm][r]{\scriptscriptstyle \pm 0.0001}}}$&$\substack{\mathmakebox[.7cm][r]{0.889}\\{\mathmakebox[.7cm][r]{\scriptscriptstyle \pm 0.0001}}}$&$\substack{\mathmakebox[.7cm][r]{84.724}\\{\mathmakebox[.7cm][r]{\scriptscriptstyle \pm 0.0645}}}$\\
\addlinespace
scopes&$\substack{\mathmakebox[.7cm][r]{133.487}\\{\mathmakebox[.7cm][r]{\scriptscriptstyle \pm 0.0493}}}$&$\substack{\mathmakebox[.7cm][r]{15.023}\\{\mathmakebox[.7cm][r]{\scriptscriptstyle \pm 0.0018}}}$&$\substack{\mathmakebox[.7cm][r]{4.511}\\{\mathmakebox[.7cm][r]{\scriptscriptstyle \pm 0.0025}}}$&$\substack{\mathmakebox[.7cm][r]{1.000}\\{\mathmakebox[.7cm][r]{\scriptscriptstyle }}}$&$\substack{\mathmakebox[.7cm][r]{0.929}\\{\mathmakebox[.7cm][r]{\scriptscriptstyle \pm 0.0005}}}$&$\substack{\mathmakebox[.7cm][r]{4.495}\\{\mathmakebox[.7cm][r]{\scriptscriptstyle \pm 0.0013}}}$&$\substack{\mathmakebox[.7cm][r]{1.000}\\{\mathmakebox[.7cm][r]{\scriptscriptstyle \pm 0.0001}}}$&$\substack{\mathmakebox[.7cm][r]{152.608}\\{\mathmakebox[.7cm][r]{\scriptscriptstyle \pm 0.0131}}}$\\
\addlinespace
smallfunc&$\substack{\mathmakebox[.7cm][r]{187.132}\\{\mathmakebox[.7cm][r]{\scriptscriptstyle \pm 0.1488}}}$&$\substack{\mathmakebox[.7cm][r]{13.078}\\{\mathmakebox[.7cm][r]{\scriptscriptstyle \pm 0.0010}}}$&$\substack{\mathmakebox[.7cm][r]{1.000}\\{\mathmakebox[.7cm][r]{\scriptscriptstyle \pm 0.0001}}}$&$\substack{\mathmakebox[.7cm][r]{1.000}\\{\mathmakebox[.7cm][r]{\scriptscriptstyle }}}$&$\substack{\mathmakebox[.7cm][r]{0.750}\\{\mathmakebox[.7cm][r]{\scriptscriptstyle \pm 0.0000}}}$&$\substack{\mathmakebox[.7cm][r]{1.000}\\{\mathmakebox[.7cm][r]{\scriptscriptstyle \pm 0.0001}}}$&$\substack{\mathmakebox[.7cm][r]{1.000}\\{\mathmakebox[.7cm][r]{\scriptscriptstyle \pm 0.0001}}}$&$\substack{\mathmakebox[.7cm][r]{230.818}\\{\mathmakebox[.7cm][r]{\scriptscriptstyle \pm 0.0145}}}$\\
\addlinespace
sum&$\substack{\mathmakebox[.7cm][r]{317.479}\\{\mathmakebox[.7cm][r]{\scriptscriptstyle \pm 0.2718}}}$&$\substack{\mathmakebox[.7cm][r]{19.362}\\{\mathmakebox[.7cm][r]{\scriptscriptstyle \pm 0.0014}}}$&$\substack{\mathmakebox[.7cm][r]{0.999}\\{\mathmakebox[.7cm][r]{\scriptscriptstyle \pm 0.0001}}}$&$\substack{\mathmakebox[.7cm][r]{1.000}\\{\mathmakebox[.7cm][r]{\scriptscriptstyle }}}$&$\substack{\mathmakebox[.7cm][r]{0.750}\\{\mathmakebox[.7cm][r]{\scriptscriptstyle \pm 0.0001}}}$&$\substack{\mathmakebox[.7cm][r]{1.000}\\{\mathmakebox[.7cm][r]{\scriptscriptstyle \pm 0.0003}}}$&$\substack{\mathmakebox[.7cm][r]{0.874}\\{\mathmakebox[.7cm][r]{\scriptscriptstyle \pm 0.0001}}}$&$\substack{\mathmakebox[.7cm][r]{418.485}\\{\mathmakebox[.7cm][r]{\scriptscriptstyle \pm 0.0921}}}$\\
\addlinespace
sum\_meth&$\substack{\mathmakebox[.7cm][r]{341.850}\\{\mathmakebox[.7cm][r]{\scriptscriptstyle \pm 1.3274}}}$&$\substack{\mathmakebox[.7cm][r]{24.106}\\{\mathmakebox[.7cm][r]{\scriptscriptstyle \pm 0.0280}}}$&$\substack{\mathmakebox[.7cm][r]{0.999}\\{\mathmakebox[.7cm][r]{\scriptscriptstyle \pm 0.0001}}}$&$\substack{\mathmakebox[.7cm][r]{1.000}\\{\mathmakebox[.7cm][r]{\scriptscriptstyle }}}$&&$\substack{\mathmakebox[.7cm][r]{1.000}\\{\mathmakebox[.7cm][r]{\scriptscriptstyle \pm 0.0001}}}$&$\substack{\mathmakebox[.7cm][r]{0.874}\\{\mathmakebox[.7cm][r]{\scriptscriptstyle \pm 0.0002}}}$&$\substack{\mathmakebox[.7cm][r]{447.472}\\{\mathmakebox[.7cm][r]{\scriptscriptstyle \pm 0.4913}}}$\\
\addlinespace
sum\_meth\_attr&$\substack{\mathmakebox[.7cm][r]{131.469}\\{\mathmakebox[.7cm][r]{\scriptscriptstyle \pm 0.7392}}}$&$\substack{\mathmakebox[.7cm][r]{17.915}\\{\mathmakebox[.7cm][r]{\scriptscriptstyle \pm 0.1052}}}$&$\substack{\mathmakebox[.7cm][r]{0.999}\\{\mathmakebox[.7cm][r]{\scriptscriptstyle \pm 0.0061}}}$&$\substack{\mathmakebox[.7cm][r]{1.000}\\{\mathmakebox[.7cm][r]{\scriptscriptstyle }}}$&&$\substack{\mathmakebox[.7cm][r]{1.131}\\{\mathmakebox[.7cm][r]{\scriptscriptstyle \pm 0.0065}}}$&$\substack{\mathmakebox[.7cm][r]{0.904}\\{\mathmakebox[.7cm][r]{\scriptscriptstyle \pm 0.0057}}}$&$\substack{\mathmakebox[.7cm][r]{145.365}\\{\mathmakebox[.7cm][r]{\scriptscriptstyle \pm 0.8321}}}$\\
\addlinespace
total\_list&$\substack{\mathmakebox[.7cm][r]{19.230}\\{\mathmakebox[.7cm][r]{\scriptscriptstyle \pm 0.0145}}}$&$\substack{\mathmakebox[.7cm][r]{2.245}\\{\mathmakebox[.7cm][r]{\scriptscriptstyle \pm 0.0008}}}$&$\substack{\mathmakebox[.7cm][r]{0.864}\\{\mathmakebox[.7cm][r]{\scriptscriptstyle \pm 0.0002}}}$&$\substack{\mathmakebox[.7cm][r]{1.000}\\{\mathmakebox[.7cm][r]{\scriptscriptstyle }}}$&$\substack{\mathmakebox[.7cm][r]{1.508}\\{\mathmakebox[.7cm][r]{\scriptscriptstyle \pm 0.0004}}}$&$\substack{\mathmakebox[.7cm][r]{0.858}\\{\mathmakebox[.7cm][r]{\scriptscriptstyle \pm 0.0005}}}$&$\substack{\mathmakebox[.7cm][r]{0.587}\\{\mathmakebox[.7cm][r]{\scriptscriptstyle \pm 0.0003}}}$&$\substack{\mathmakebox[.7cm][r]{33.667}\\{\mathmakebox[.7cm][r]{\scriptscriptstyle \pm 0.0633}}}$\\
\addlinespace
walk\_list&$\substack{\mathmakebox[.7cm][r]{5.060}\\{\mathmakebox[.7cm][r]{\scriptscriptstyle \pm 0.0071}}}$&$\substack{\mathmakebox[.7cm][r]{0.406}\\{\mathmakebox[.7cm][r]{\scriptscriptstyle \pm 0.0005}}}$&$\substack{\mathmakebox[.7cm][r]{0.779}\\{\mathmakebox[.7cm][r]{\scriptscriptstyle \pm 0.0011}}}$&$\substack{\mathmakebox[.7cm][r]{1.000}\\{\mathmakebox[.7cm][r]{\scriptscriptstyle }}}$&$\substack{\mathmakebox[.7cm][r]{1.601}\\{\mathmakebox[.7cm][r]{\scriptscriptstyle \pm 0.0026}}}$&$\substack{\mathmakebox[.7cm][r]{1.010}\\{\mathmakebox[.7cm][r]{\scriptscriptstyle \pm 0.0018}}}$&$\substack{\mathmakebox[.7cm][r]{1.080}\\{\mathmakebox[.7cm][r]{\scriptscriptstyle \pm 0.0015}}}$&$\substack{\mathmakebox[.7cm][r]{10.647}\\{\mathmakebox[.7cm][r]{\scriptscriptstyle \pm 0.0677}}}$\\
\midrule
deltablue&$\substack{\mathmakebox[.7cm][r]{16.528}\\{\mathmakebox[.7cm][r]{\scriptscriptstyle \pm 0.0707}}}$&$\substack{\mathmakebox[.7cm][r]{671.482}\\{\mathmakebox[.7cm][r]{\scriptscriptstyle \pm 2.9041}}}$&$\substack{\mathmakebox[.7cm][r]{4.325}\\{\mathmakebox[.7cm][r]{\scriptscriptstyle \pm 0.0212}}}$&$\substack{\mathmakebox[.7cm][r]{1.000}\\{\mathmakebox[.7cm][r]{\scriptscriptstyle }}}$&&$\substack{\mathmakebox[.7cm][r]{4.507}\\{\mathmakebox[.7cm][r]{\scriptscriptstyle \pm 0.0214}}}$&$\substack{\mathmakebox[.7cm][r]{0.457}\\{\mathmakebox[.7cm][r]{\scriptscriptstyle \pm 0.0026}}}$&$\substack{\mathmakebox[.7cm][r]{144.149}\\{\mathmakebox[.7cm][r]{\scriptscriptstyle \pm 2.6843}}}$\\
\addlinespace
fannkuch&$\substack{\mathmakebox[.7cm][r]{20.582}\\{\mathmakebox[.7cm][r]{\scriptscriptstyle \pm 0.0226}}}$&$\substack{\mathmakebox[.7cm][r]{3.342}\\{\mathmakebox[.7cm][r]{\scriptscriptstyle \pm 0.0025}}}$&$\substack{\mathmakebox[.7cm][r]{1.848}\\{\mathmakebox[.7cm][r]{\scriptscriptstyle \pm 0.0007}}}$&$\substack{\mathmakebox[.7cm][r]{1.000}\\{\mathmakebox[.7cm][r]{\scriptscriptstyle }}}$&$\substack{\mathmakebox[.7cm][r]{1.891}\\{\mathmakebox[.7cm][r]{\scriptscriptstyle \pm 0.0005}}}$&$\substack{\mathmakebox[.7cm][r]{1.878}\\{\mathmakebox[.7cm][r]{\scriptscriptstyle \pm 0.0005}}}$&$\substack{\mathmakebox[.7cm][r]{1.005}\\{\mathmakebox[.7cm][r]{\scriptscriptstyle \pm 0.0004}}}$&$\substack{\mathmakebox[.7cm][r]{14.387}\\{\mathmakebox[.7cm][r]{\scriptscriptstyle \pm 0.0128}}}$\\
\addlinespace
mandel&&$\substack{\mathmakebox[.7cm][r]{0.791}\\{\mathmakebox[.7cm][r]{\scriptscriptstyle \pm 0.0056}}}$&$\substack{\mathmakebox[.7cm][r]{0.921}\\{\mathmakebox[.7cm][r]{\scriptscriptstyle \pm 0.0005}}}$&$\substack{\mathmakebox[.7cm][r]{1.000}\\{\mathmakebox[.7cm][r]{\scriptscriptstyle }}}$&$\substack{\mathmakebox[.7cm][r]{0.493}\\{\mathmakebox[.7cm][r]{\scriptscriptstyle \pm 0.0001}}}$&$\substack{\mathmakebox[.7cm][r]{0.999}\\{\mathmakebox[.7cm][r]{\scriptscriptstyle \pm 0.0003}}}$&&$\substack{\mathmakebox[.7cm][r]{7.241}\\{\mathmakebox[.7cm][r]{\scriptscriptstyle \pm 0.0188}}}$\\
\addlinespace
richards&$\substack{\mathmakebox[.7cm][r]{26.902}\\{\mathmakebox[.7cm][r]{\scriptscriptstyle \pm 0.0189}}}$&$\substack{\mathmakebox[.7cm][r]{11.897}\\{\mathmakebox[.7cm][r]{\scriptscriptstyle \pm 0.0088}}}$&$\substack{\mathmakebox[.7cm][r]{0.853}\\{\mathmakebox[.7cm][r]{\scriptscriptstyle \pm 0.0010}}}$&$\substack{\mathmakebox[.7cm][r]{1.000}\\{\mathmakebox[.7cm][r]{\scriptscriptstyle }}}$&&$\substack{\mathmakebox[.7cm][r]{0.887}\\{\mathmakebox[.7cm][r]{\scriptscriptstyle \pm 0.0007}}}$&$\substack{\mathmakebox[.7cm][r]{0.488}\\{\mathmakebox[.7cm][r]{\scriptscriptstyle \pm 0.0005}}}$&$\substack{\mathmakebox[.7cm][r]{24.207}\\{\mathmakebox[.7cm][r]{\scriptscriptstyle \pm 0.0236}}}$\\
\midrule
Geometric Mean&$\substack{\mathmakebox[.7cm][r]{52.743}\\{\mathmakebox[.7cm][r]{\scriptscriptstyle \pm 0.0341}}}$&$\substack{\mathmakebox[.7cm][r]{6.940}\\{\mathmakebox[.7cm][r]{\scriptscriptstyle \pm 0.0047}}}$&$\substack{\mathmakebox[.7cm][r]{1.222}\\{\mathmakebox[.7cm][r]{\scriptscriptstyle \pm 0.0006}}}$&$\substack{\mathmakebox[.7cm][r]{1.000}\\{\mathmakebox[.7cm][r]{\scriptscriptstyle }}}$&$\substack{\mathmakebox[.7cm][r]{0.963}\\{\mathmakebox[.7cm][r]{\scriptscriptstyle \pm 0.0003}}}$&$\substack{\mathmakebox[.7cm][r]{1.277}\\{\mathmakebox[.7cm][r]{\scriptscriptstyle \pm 0.0006}}}$&$\substack{\mathmakebox[.7cm][r]{0.813}\\{\mathmakebox[.7cm][r]{\scriptscriptstyle \pm 0.0007}}}$&$\substack{\mathmakebox[.7cm][r]{55.549}\\{\mathmakebox[.7cm][r]{\scriptscriptstyle \pm 0.0692}}}$\\
\bottomrule
\end{tabular}
\end{adjustbox}
\vspace{0.5em}
\caption{Microbenchmark timings relative to \ourvmcomp. Note that \ourvmcomp
and \ourvmcompr are the same VM, but running composed-PHP and composed-Python
benchmark variants respectively.}
\label{tab:relresults}
\end{table*}
Table~\ref{tab:relresults} shows the results of our microbenchmarks relative to the
composed \ourvm variant.
Absolute timings are shown in Table~\ref{tab:absresults}
in the Appendix. Starting with the simplest observations, we can
see that Zend and CPython (C-based interpreters) are slower than HHVM, \hippy,
and \pypy (JIT-based VMs). Small to medium benchmarks tend to flatter
meta-tracing, and so \hippy and \pypy outperform HHVM. \pypy is nearly always
faster than \hippy, reflecting the greater level of engineering \pypy has
received.

\ourvmmono's results are very similar to \hippy's, though a few cases run slightly
slower on \ourvm. For walk\_list and DeltaBlue, some missing optimisations
in \ourvm's scoping lookups cause undue bloat in the optimised traces.
instchain and sum\_meth\_attr in contrast have identical traces except
for a small portion of their headers: this seemingly small
difference has a surprisingly large run-time effect which we do not
fully understand.

\ourvm is generally faster than \hippy on the composed-PHP benchmarks. This is largely due to moving code
from PHP (slower \hippy) to Python (faster \pypy) and the ability that
meta-tracing has to naturally inline code across both languages.  \ourvm is
in most cases slower than \pypy, as we would expect, because of the additional
overhead of adapters and cross-language scoping. Although meta-tracing naturally
optimises the vast majority of these operations away, a few inevitably remain,
and their cumulative effect is often noticeable, even though it is not severe.
On the geometric mean \ourvmcomp is only around 20\%{} slower on average than
\pypy, and no individual \ourvmcomp benchmark is more than 2.2x slower. The
composed-Python benchmarks have a similar overall average to the composed-PHP
benchmarks, though several benchmarks are slower. By comparing traces
from the composed-PHP and composed-Python benchmarks, we were able to
identify several missing optimisations in \hippy that are likely to
account for most such slowdowns: redundant
comparisons in logical operators; many more allocations in PHP iterators; and
more allocations when appending to PHP lists.

In some cases, the timings for composed variants running on \ourvm 
are virtually identical to the mono-language variants running on \ourvm's constituents (e.g.
smallfunc on \hippy, \ourvmcomp and \pypy are all roughly 1x).
For small benchmarks, we would expect any well-written RPython VM
to compile virtually identical traces, and such benchmarks
show this effect. However, in some cases where we expected identical
performance for both \ourvm and its constituents, the composed
variant is faster: lla0r, l1a1r, and smallfunc. Indeed, the crucial
parts of the traces were identical between the two VMs in all these cases. Further exploration showed
that RPython's machine code generator occasionally emits less than optimal code
(in this case unnecessary x86-64 \texttt{MOV}s) that account for the difference.
We do not understand the precise reason for this, but it seems plausible it is a
limitation of the current register allocator. We have reported our findings to
the RPython developers.

\addtolength{\tabcolsep}{-.4em}
\begin{table*}
\small
\centering

\begin{adjustbox}{width=1.2\textwidth,center}
\begin{tabular}{rrrp{.1em}rrrp{.1em}rrrp{.1em}rrrp{.1em}rrr}
\toprule
{\tiny \color{gray}{p$_{1}$}:}&$\substack{\mathmakebox[.7cm][r]{\color{gray}{0.246}}\color{gray}{s}\\{\mathmakebox[.7cm][r]{\scriptscriptstyle \color{gray}{\pm 0.0005}}}}$&$\substack{\mathmakebox[.7cm][r]{\color{gray}{1.000}}\color{gray}{\times}\\{\mathmakebox[.7cm][r]{\scriptscriptstyle \color{gray}{\pm 0.0029}}}}$&&{\tiny \color{gray}{p$_{17}$}:}&$\substack{\mathmakebox[.7cm][r]{\color{gray}{0.246}}\color{gray}{s}\\{\mathmakebox[.7cm][r]{\scriptscriptstyle \color{gray}{\pm 0.0004}}}}$&$\substack{\mathmakebox[.7cm][r]{\color{gray}{1.000}}\color{gray}{\times}\\{\mathmakebox[.7cm][r]{\scriptscriptstyle \color{gray}{\pm 0.0028}}}}$&&{\tiny \color{gray}{p$_{33}$}:}&$\substack{\mathmakebox[.7cm][r]{\color{gray}{0.246}}\color{gray}{s}\\{\mathmakebox[.7cm][r]{\scriptscriptstyle \color{gray}{\pm 0.0005}}}}$&$\substack{\mathmakebox[.7cm][r]{\color{gray}{0.999}}\color{gray}{\times}\\{\mathmakebox[.7cm][r]{\scriptscriptstyle \color{gray}{\pm 0.0028}}}}$&&{\tiny \color{gray}{p$_{49}$}:}&$\substack{\mathmakebox[.7cm][r]{\color{gray}{0.246}}\color{gray}{s}\\{\mathmakebox[.7cm][r]{\scriptscriptstyle \color{gray}{\pm 0.0005}}}}$&$\substack{\mathmakebox[.7cm][r]{\color{gray}{1.002}}\color{gray}{\times}\\{\mathmakebox[.7cm][r]{\scriptscriptstyle \color{gray}{\pm 0.0028}}}}$&&{\tiny \color{gray}{p$_{65}$}:}&$\substack{\mathmakebox[.7cm][r]{\color{gray}{0.246}}\color{gray}{s}\\{\mathmakebox[.7cm][r]{\scriptscriptstyle \color{gray}{\pm 0.0005}}}}$&$\substack{\mathmakebox[.7cm][r]{\color{gray}{1.001}}\color{gray}{\times}\\{\mathmakebox[.7cm][r]{\scriptscriptstyle \color{gray}{\pm 0.0029}}}}$\\
\cellcolor{black!5}{\tiny \textbf{p$_{2}$}:}&\cellcolor{black!5}$\substack{\mathmakebox[.7cm][r]{\mathbf{0.120}}\mathbf{s}\\{\mathmakebox[.7cm][r]{\scriptscriptstyle \mathbf{\pm 0.0003}}}}$&\cellcolor{black!5}$\substack{\mathmakebox[.7cm][r]{\mathbf{0.490}}\mathbf{\times}\\{\mathmakebox[.7cm][r]{\scriptscriptstyle \mathbf{\pm 0.0015}}}}$&&\cellcolor{black!5}{\tiny p$_{18}$:}&\cellcolor{black!5}$\substack{\mathmakebox[.7cm][r]{0.250}s\\{\mathmakebox[.7cm][r]{\scriptscriptstyle \pm 0.0005}}}$&\cellcolor{black!5}$\substack{\mathmakebox[.7cm][r]{1.015}\times\\{\mathmakebox[.7cm][r]{\scriptscriptstyle \pm 0.0029}}}$&&\cellcolor{black!5}{\tiny \color{gray}{p$_{34}$}:}&\cellcolor{black!5}$\substack{\mathmakebox[.7cm][r]{\color{gray}{0.246}}\color{gray}{s}\\{\mathmakebox[.7cm][r]{\scriptscriptstyle \color{gray}{\pm 0.0004}}}}$&\cellcolor{black!5}$\substack{\mathmakebox[.7cm][r]{\color{gray}{1.000}}\color{gray}{\times}\\{\mathmakebox[.7cm][r]{\scriptscriptstyle \color{gray}{\pm 0.0026}}}}$&&\cellcolor{black!5}{\tiny \color{gray}{p$_{50}$}:}&\cellcolor{black!5}$\substack{\mathmakebox[.7cm][r]{\color{gray}{0.246}}\color{gray}{s}\\{\mathmakebox[.7cm][r]{\scriptscriptstyle \color{gray}{\pm 0.0006}}}}$&\cellcolor{black!5}$\substack{\mathmakebox[.7cm][r]{\color{gray}{1.001}}\color{gray}{\times}\\{\mathmakebox[.7cm][r]{\scriptscriptstyle \color{gray}{\pm 0.0032}}}}$&&\cellcolor{black!5}{\tiny p$_{66}$:}&\cellcolor{black!5}$\substack{\mathmakebox[.7cm][r]{0.247}s\\{\mathmakebox[.7cm][r]{\scriptscriptstyle \pm 0.0004}}}$&\cellcolor{black!5}$\substack{\mathmakebox[.7cm][r]{1.006}\times\\{\mathmakebox[.7cm][r]{\scriptscriptstyle \pm 0.0026}}}$\\
{\tiny p$_{3}$:}&$\substack{\mathmakebox[.7cm][r]{0.240}s\\{\mathmakebox[.7cm][r]{\scriptscriptstyle \pm 0.0004}}}$&$\substack{\mathmakebox[.7cm][r]{0.978}\times\\{\mathmakebox[.7cm][r]{\scriptscriptstyle \pm 0.0026}}}$&&{\tiny p$_{19}$:}&$\substack{\mathmakebox[.7cm][r]{0.245}s\\{\mathmakebox[.7cm][r]{\scriptscriptstyle \pm 0.0004}}}$&$\substack{\mathmakebox[.7cm][r]{0.999}\times\\{\mathmakebox[.7cm][r]{\scriptscriptstyle \pm 0.0026}}}$&&{\tiny \color{gray}{p$_{35}$}:}&$\substack{\mathmakebox[.7cm][r]{\color{gray}{0.246}}\color{gray}{s}\\{\mathmakebox[.7cm][r]{\scriptscriptstyle \color{gray}{\pm 0.0006}}}}$&$\substack{\mathmakebox[.7cm][r]{\color{gray}{1.000}}\color{gray}{\times}\\{\mathmakebox[.7cm][r]{\scriptscriptstyle \color{gray}{\pm 0.0031}}}}$&&{\tiny \color{gray}{p$_{51}$}:}&$\substack{\mathmakebox[.7cm][r]{\color{gray}{0.246}}\color{gray}{s}\\{\mathmakebox[.7cm][r]{\scriptscriptstyle \color{gray}{\pm 0.0004}}}}$&$\substack{\mathmakebox[.7cm][r]{\color{gray}{1.000}}\color{gray}{\times}\\{\mathmakebox[.7cm][r]{\scriptscriptstyle \color{gray}{\pm 0.0027}}}}$&&{\tiny \color{gray}{p$_{67}$}:}&$\substack{\mathmakebox[.7cm][r]{\color{gray}{0.246}}\color{gray}{s}\\{\mathmakebox[.7cm][r]{\scriptscriptstyle \color{gray}{\pm 0.0005}}}}$&$\substack{\mathmakebox[.7cm][r]{\color{gray}{1.001}}\color{gray}{\times}\\{\mathmakebox[.7cm][r]{\scriptscriptstyle \color{gray}{\pm 0.0027}}}}$\\
\cellcolor{black!5}{\tiny p$_{4}$:}&\cellcolor{black!5}$\substack{\mathmakebox[.7cm][r]{0.249}s\\{\mathmakebox[.7cm][r]{\scriptscriptstyle \pm 0.0005}}}$&\cellcolor{black!5}$\substack{\mathmakebox[.7cm][r]{1.015}\times\\{\mathmakebox[.7cm][r]{\scriptscriptstyle \pm 0.0030}}}$&&\cellcolor{black!5}{\tiny p$_{20}$:}&\cellcolor{black!5}$\substack{\mathmakebox[.7cm][r]{0.245}s\\{\mathmakebox[.7cm][r]{\scriptscriptstyle \pm 0.0004}}}$&\cellcolor{black!5}$\substack{\mathmakebox[.7cm][r]{0.998}\times\\{\mathmakebox[.7cm][r]{\scriptscriptstyle \pm 0.0026}}}$&&\cellcolor{black!5}{\tiny \color{gray}{p$_{36}$}:}&\cellcolor{black!5}$\substack{\mathmakebox[.7cm][r]{\color{gray}{0.246}}\color{gray}{s}\\{\mathmakebox[.7cm][r]{\scriptscriptstyle \color{gray}{\pm 0.0005}}}}$&\cellcolor{black!5}$\substack{\mathmakebox[.7cm][r]{\color{gray}{1.002}}\color{gray}{\times}\\{\mathmakebox[.7cm][r]{\scriptscriptstyle \color{gray}{\pm 0.0030}}}}$&&\cellcolor{black!5}{\tiny \color{gray}{p$_{52}$}:}&\cellcolor{black!5}$\substack{\mathmakebox[.7cm][r]{\color{gray}{0.246}}\color{gray}{s}\\{\mathmakebox[.7cm][r]{\scriptscriptstyle \color{gray}{\pm 0.0004}}}}$&\cellcolor{black!5}$\substack{\mathmakebox[.7cm][r]{\color{gray}{1.000}}\color{gray}{\times}\\{\mathmakebox[.7cm][r]{\scriptscriptstyle \color{gray}{\pm 0.0026}}}}$&&\cellcolor{black!5}{\tiny \color{gray}{p$_{68}$}:}&\cellcolor{black!5}$\substack{\mathmakebox[.7cm][r]{\color{gray}{0.246}}\color{gray}{s}\\{\mathmakebox[.7cm][r]{\scriptscriptstyle \color{gray}{\pm 0.0005}}}}$&\cellcolor{black!5}$\substack{\mathmakebox[.7cm][r]{\color{gray}{1.000}}\color{gray}{\times}\\{\mathmakebox[.7cm][r]{\scriptscriptstyle \color{gray}{\pm 0.0030}}}}$\\
{\tiny \textbf{p$_{5}$}:}&$\substack{\mathmakebox[.7cm][r]{\mathbf{0.132}}\mathbf{s}\\{\mathmakebox[.7cm][r]{\scriptscriptstyle \mathbf{\pm 0.0002}}}}$&$\substack{\mathmakebox[.7cm][r]{\mathbf{0.538}}\mathbf{\times}\\{\mathmakebox[.7cm][r]{\scriptscriptstyle \mathbf{\pm 0.0014}}}}$&&{\tiny \color{gray}{p$_{21}$}:}&$\substack{\mathmakebox[.7cm][r]{\color{gray}{0.246}}\color{gray}{s}\\{\mathmakebox[.7cm][r]{\scriptscriptstyle \color{gray}{\pm 0.0004}}}}$&$\substack{\mathmakebox[.7cm][r]{\color{gray}{1.001}}\color{gray}{\times}\\{\mathmakebox[.7cm][r]{\scriptscriptstyle \color{gray}{\pm 0.0027}}}}$&&{\tiny \color{gray}{p$_{37}$}:}&$\substack{\mathmakebox[.7cm][r]{\color{gray}{0.246}}\color{gray}{s}\\{\mathmakebox[.7cm][r]{\scriptscriptstyle \color{gray}{\pm 0.0005}}}}$&$\substack{\mathmakebox[.7cm][r]{\color{gray}{1.001}}\color{gray}{\times}\\{\mathmakebox[.7cm][r]{\scriptscriptstyle \color{gray}{\pm 0.0028}}}}$&&{\tiny p$_{53}$:}&$\substack{\mathmakebox[.7cm][r]{0.248}s\\{\mathmakebox[.7cm][r]{\scriptscriptstyle \pm 0.0004}}}$&$\substack{\mathmakebox[.7cm][r]{1.008}\times\\{\mathmakebox[.7cm][r]{\scriptscriptstyle \pm 0.0027}}}$&&{\tiny p$_{69}$:}&$\substack{\mathmakebox[.7cm][r]{0.251}s\\{\mathmakebox[.7cm][r]{\scriptscriptstyle \pm 0.0004}}}$&$\substack{\mathmakebox[.7cm][r]{1.021}\times\\{\mathmakebox[.7cm][r]{\scriptscriptstyle \pm 0.0028}}}$\\
\cellcolor{black!5}{\tiny \textbf{p$_{6}$}:}&\cellcolor{black!5}$\substack{\mathmakebox[.7cm][r]{\mathbf{0.131}}\mathbf{s}\\{\mathmakebox[.7cm][r]{\scriptscriptstyle \mathbf{\pm 0.0002}}}}$&\cellcolor{black!5}$\substack{\mathmakebox[.7cm][r]{\mathbf{0.533}}\mathbf{\times}\\{\mathmakebox[.7cm][r]{\scriptscriptstyle \mathbf{\pm 0.0013}}}}$&&\cellcolor{black!5}{\tiny \color{gray}{p$_{22}$}:}&\cellcolor{black!5}$\substack{\mathmakebox[.7cm][r]{\color{gray}{0.247}}\color{gray}{s}\\{\mathmakebox[.7cm][r]{\scriptscriptstyle \color{gray}{\pm 0.0005}}}}$&\cellcolor{black!5}$\substack{\mathmakebox[.7cm][r]{\color{gray}{1.005}}\color{gray}{\times}\\{\mathmakebox[.7cm][r]{\scriptscriptstyle \color{gray}{\pm 0.0029}}}}$&&\cellcolor{black!5}{\tiny \color{gray}{p$_{38}$}:}&\cellcolor{black!5}$\substack{\mathmakebox[.7cm][r]{\color{gray}{0.246}}\color{gray}{s}\\{\mathmakebox[.7cm][r]{\scriptscriptstyle \color{gray}{\pm 0.0005}}}}$&\cellcolor{black!5}$\substack{\mathmakebox[.7cm][r]{\color{gray}{1.000}}\color{gray}{\times}\\{\mathmakebox[.7cm][r]{\scriptscriptstyle \color{gray}{\pm 0.0028}}}}$&&\cellcolor{black!5}{\tiny \color{gray}{p$_{54}$}:}&\cellcolor{black!5}$\substack{\mathmakebox[.7cm][r]{\color{gray}{0.246}}\color{gray}{s}\\{\mathmakebox[.7cm][r]{\scriptscriptstyle \color{gray}{\pm 0.0005}}}}$&\cellcolor{black!5}$\substack{\mathmakebox[.7cm][r]{\color{gray}{0.999}}\color{gray}{\times}\\{\mathmakebox[.7cm][r]{\scriptscriptstyle \color{gray}{\pm 0.0028}}}}$&&\cellcolor{black!5}{\tiny p$_{70}$:}&\cellcolor{black!5}$\substack{\mathmakebox[.7cm][r]{0.248}s\\{\mathmakebox[.7cm][r]{\scriptscriptstyle \pm 0.0005}}}$&\cellcolor{black!5}$\substack{\mathmakebox[.7cm][r]{1.008}\times\\{\mathmakebox[.7cm][r]{\scriptscriptstyle \pm 0.0028}}}$\\
{\tiny \textbf{p$_{7}$}:}&$\substack{\mathmakebox[.7cm][r]{\mathbf{0.175}}\mathbf{s}\\{\mathmakebox[.7cm][r]{\scriptscriptstyle \mathbf{\pm 0.0003}}}}$&$\substack{\mathmakebox[.7cm][r]{\mathbf{0.710}}\mathbf{\times}\\{\mathmakebox[.7cm][r]{\scriptscriptstyle \mathbf{\pm 0.0020}}}}$&&{\tiny p$_{23}$:}&$\substack{\mathmakebox[.7cm][r]{0.244}s\\{\mathmakebox[.7cm][r]{\scriptscriptstyle \pm 0.0005}}}$&$\substack{\mathmakebox[.7cm][r]{0.993}\times\\{\mathmakebox[.7cm][r]{\scriptscriptstyle \pm 0.0027}}}$&&{\tiny \color{gray}{p$_{39}$}:}&$\substack{\mathmakebox[.7cm][r]{\color{gray}{0.246}}\color{gray}{s}\\{\mathmakebox[.7cm][r]{\scriptscriptstyle \color{gray}{\pm 0.0005}}}}$&$\substack{\mathmakebox[.7cm][r]{\color{gray}{0.999}}\color{gray}{\times}\\{\mathmakebox[.7cm][r]{\scriptscriptstyle \color{gray}{\pm 0.0027}}}}$&&{\tiny \color{gray}{p$_{55}$}:}&$\substack{\mathmakebox[.7cm][r]{\color{gray}{0.247}}\color{gray}{s}\\{\mathmakebox[.7cm][r]{\scriptscriptstyle \color{gray}{\pm 0.0005}}}}$&$\substack{\mathmakebox[.7cm][r]{\color{gray}{1.004}}\color{gray}{\times}\\{\mathmakebox[.7cm][r]{\scriptscriptstyle \color{gray}{\pm 0.0028}}}}$&&{\tiny p$_{71}$:}&$\substack{\mathmakebox[.7cm][r]{0.242}s\\{\mathmakebox[.7cm][r]{\scriptscriptstyle \pm 0.0004}}}$&$\substack{\mathmakebox[.7cm][r]{0.986}\times\\{\mathmakebox[.7cm][r]{\scriptscriptstyle \pm 0.0025}}}$\\
\cellcolor{black!5}{\tiny \color{gray}{p$_{8}$}:}&\cellcolor{black!5}$\substack{\mathmakebox[.7cm][r]{\color{gray}{0.246}}\color{gray}{s}\\{\mathmakebox[.7cm][r]{\scriptscriptstyle \color{gray}{\pm 0.0006}}}}$&\cellcolor{black!5}$\substack{\mathmakebox[.7cm][r]{\color{gray}{1.002}}\color{gray}{\times}\\{\mathmakebox[.7cm][r]{\scriptscriptstyle \color{gray}{\pm 0.0032}}}}$&&\cellcolor{black!5}{\tiny \color{gray}{p$_{24}$}:}&\cellcolor{black!5}$\substack{\mathmakebox[.7cm][r]{\color{gray}{0.246}}\color{gray}{s}\\{\mathmakebox[.7cm][r]{\scriptscriptstyle \color{gray}{\pm 0.0005}}}}$&\cellcolor{black!5}$\substack{\mathmakebox[.7cm][r]{\color{gray}{1.000}}\color{gray}{\times}\\{\mathmakebox[.7cm][r]{\scriptscriptstyle \color{gray}{\pm 0.0029}}}}$&&\cellcolor{black!5}{\tiny \color{gray}{p$_{40}$}:}&\cellcolor{black!5}$\substack{\mathmakebox[.7cm][r]{\color{gray}{0.246}}\color{gray}{s}\\{\mathmakebox[.7cm][r]{\scriptscriptstyle \color{gray}{\pm 0.0005}}}}$&\cellcolor{black!5}$\substack{\mathmakebox[.7cm][r]{\color{gray}{1.000}}\color{gray}{\times}\\{\mathmakebox[.7cm][r]{\scriptscriptstyle \color{gray}{\pm 0.0030}}}}$&&\cellcolor{black!5}{\tiny \color{gray}{p$_{56}$}:}&\cellcolor{black!5}$\substack{\mathmakebox[.7cm][r]{\color{gray}{0.247}}\color{gray}{s}\\{\mathmakebox[.7cm][r]{\scriptscriptstyle \color{gray}{\pm 0.0006}}}}$&\cellcolor{black!5}$\substack{\mathmakebox[.7cm][r]{\color{gray}{1.005}}\color{gray}{\times}\\{\mathmakebox[.7cm][r]{\scriptscriptstyle \color{gray}{\pm 0.0031}}}}$&&\cellcolor{black!5}{\tiny \color{gray}{p$_{72}$}:}&\cellcolor{black!5}$\substack{\mathmakebox[.7cm][r]{\color{gray}{0.246}}\color{gray}{s}\\{\mathmakebox[.7cm][r]{\scriptscriptstyle \color{gray}{\pm 0.0005}}}}$&\cellcolor{black!5}$\substack{\mathmakebox[.7cm][r]{\color{gray}{1.001}}\color{gray}{\times}\\{\mathmakebox[.7cm][r]{\scriptscriptstyle \color{gray}{\pm 0.0028}}}}$\\
{\tiny \color{gray}{p$_{9}$}:}&$\substack{\mathmakebox[.7cm][r]{\color{gray}{0.246}}\color{gray}{s}\\{\mathmakebox[.7cm][r]{\scriptscriptstyle \color{gray}{\pm 0.0005}}}}$&$\substack{\mathmakebox[.7cm][r]{\color{gray}{1.000}}\color{gray}{\times}\\{\mathmakebox[.7cm][r]{\scriptscriptstyle \color{gray}{\pm 0.0029}}}}$&&{\tiny \color{gray}{p$_{25}$}:}&$\substack{\mathmakebox[.7cm][r]{\color{gray}{0.246}}\color{gray}{s}\\{\mathmakebox[.7cm][r]{\scriptscriptstyle \color{gray}{\pm 0.0005}}}}$&$\substack{\mathmakebox[.7cm][r]{\color{gray}{1.002}}\color{gray}{\times}\\{\mathmakebox[.7cm][r]{\scriptscriptstyle \color{gray}{\pm 0.0029}}}}$&&{\tiny \color{gray}{p$_{41}$}:}&$\substack{\mathmakebox[.7cm][r]{\color{gray}{0.245}}\color{gray}{s}\\{\mathmakebox[.7cm][r]{\scriptscriptstyle \color{gray}{\pm 0.0005}}}}$&$\substack{\mathmakebox[.7cm][r]{\color{gray}{0.995}}\color{gray}{\times}\\{\mathmakebox[.7cm][r]{\scriptscriptstyle \color{gray}{\pm 0.0028}}}}$&&{\tiny \color{gray}{p$_{57}$}:}&$\substack{\mathmakebox[.7cm][r]{\color{gray}{0.245}}\color{gray}{s}\\{\mathmakebox[.7cm][r]{\scriptscriptstyle \color{gray}{\pm 0.0005}}}}$&$\substack{\mathmakebox[.7cm][r]{\color{gray}{0.999}}\color{gray}{\times}\\{\mathmakebox[.7cm][r]{\scriptscriptstyle \color{gray}{\pm 0.0028}}}}$&&{\tiny \color{gray}{p$_{73}$}:}&$\substack{\mathmakebox[.7cm][r]{\color{gray}{0.246}}\color{gray}{s}\\{\mathmakebox[.7cm][r]{\scriptscriptstyle \color{gray}{\pm 0.0005}}}}$&$\substack{\mathmakebox[.7cm][r]{\color{gray}{1.000}}\color{gray}{\times}\\{\mathmakebox[.7cm][r]{\scriptscriptstyle \color{gray}{\pm 0.0027}}}}$\\
\cellcolor{black!5}{\tiny \color{gray}{p$_{10}$}:}&\cellcolor{black!5}$\substack{\mathmakebox[.7cm][r]{\color{gray}{0.246}}\color{gray}{s}\\{\mathmakebox[.7cm][r]{\scriptscriptstyle \color{gray}{\pm 0.0004}}}}$&\cellcolor{black!5}$\substack{\mathmakebox[.7cm][r]{\color{gray}{1.000}}\color{gray}{\times}\\{\mathmakebox[.7cm][r]{\scriptscriptstyle \color{gray}{\pm 0.0026}}}}$&&\cellcolor{black!5}{\tiny \color{gray}{p$_{26}$}:}&\cellcolor{black!5}$\substack{\mathmakebox[.7cm][r]{\color{gray}{0.246}}\color{gray}{s}\\{\mathmakebox[.7cm][r]{\scriptscriptstyle \color{gray}{\pm 0.0004}}}}$&\cellcolor{black!5}$\substack{\mathmakebox[.7cm][r]{\color{gray}{1.001}}\color{gray}{\times}\\{\mathmakebox[.7cm][r]{\scriptscriptstyle \color{gray}{\pm 0.0027}}}}$&&\cellcolor{black!5}{\tiny p$_{42}$:}&\cellcolor{black!5}$\substack{\mathmakebox[.7cm][r]{0.248}s\\{\mathmakebox[.7cm][r]{\scriptscriptstyle \pm 0.0004}}}$&\cellcolor{black!5}$\substack{\mathmakebox[.7cm][r]{1.011}\times\\{\mathmakebox[.7cm][r]{\scriptscriptstyle \pm 0.0027}}}$&&\cellcolor{black!5}{\tiny p$_{58}$:}&\cellcolor{black!5}$\substack{\mathmakebox[.7cm][r]{0.245}s\\{\mathmakebox[.7cm][r]{\scriptscriptstyle \pm 0.0004}}}$&\cellcolor{black!5}$\substack{\mathmakebox[.7cm][r]{0.999}\times\\{\mathmakebox[.7cm][r]{\scriptscriptstyle \pm 0.0026}}}$&&\cellcolor{black!5}{\tiny p$_{74}$:}&\cellcolor{black!5}$\substack{\mathmakebox[.7cm][r]{0.248}s\\{\mathmakebox[.7cm][r]{\scriptscriptstyle \pm 0.0004}}}$&\cellcolor{black!5}$\substack{\mathmakebox[.7cm][r]{1.011}\times\\{\mathmakebox[.7cm][r]{\scriptscriptstyle \pm 0.0027}}}$\\
{\tiny \color{gray}{p$_{11}$}:}&$\substack{\mathmakebox[.7cm][r]{\color{gray}{0.246}}\color{gray}{s}\\{\mathmakebox[.7cm][r]{\scriptscriptstyle \color{gray}{\pm 0.0005}}}}$&$\substack{\mathmakebox[.7cm][r]{\color{gray}{1.000}}\color{gray}{\times}\\{\mathmakebox[.7cm][r]{\scriptscriptstyle \color{gray}{\pm 0.0027}}}}$&&{\tiny \color{gray}{p$_{27}$}:}&$\substack{\mathmakebox[.7cm][r]{\color{gray}{0.247}}\color{gray}{s}\\{\mathmakebox[.7cm][r]{\scriptscriptstyle \color{gray}{\pm 0.0006}}}}$&$\substack{\mathmakebox[.7cm][r]{\color{gray}{1.006}}\color{gray}{\times}\\{\mathmakebox[.7cm][r]{\scriptscriptstyle \color{gray}{\pm 0.0031}}}}$&&{\tiny \color{gray}{p$_{43}$}:}&$\substack{\mathmakebox[.7cm][r]{\color{gray}{0.247}}\color{gray}{s}\\{\mathmakebox[.7cm][r]{\scriptscriptstyle \color{gray}{\pm 0.0005}}}}$&$\substack{\mathmakebox[.7cm][r]{\color{gray}{1.005}}\color{gray}{\times}\\{\mathmakebox[.7cm][r]{\scriptscriptstyle \color{gray}{\pm 0.0029}}}}$&&{\tiny p$_{59}$:}&$\substack{\mathmakebox[.7cm][r]{0.248}s\\{\mathmakebox[.7cm][r]{\scriptscriptstyle \pm 0.0006}}}$&$\substack{\mathmakebox[.7cm][r]{1.011}\times\\{\mathmakebox[.7cm][r]{\scriptscriptstyle \pm 0.0032}}}$&&{\tiny p$_{75}$:}&$\substack{\mathmakebox[.7cm][r]{0.251}s\\{\mathmakebox[.7cm][r]{\scriptscriptstyle \pm 0.0004}}}$&$\substack{\mathmakebox[.7cm][r]{1.021}\times\\{\mathmakebox[.7cm][r]{\scriptscriptstyle \pm 0.0027}}}$\\
\cellcolor{black!5}{\tiny \color{gray}{p$_{12}$}:}&\cellcolor{black!5}$\substack{\mathmakebox[.7cm][r]{\color{gray}{0.246}}\color{gray}{s}\\{\mathmakebox[.7cm][r]{\scriptscriptstyle \color{gray}{\pm 0.0004}}}}$&\cellcolor{black!5}$\substack{\mathmakebox[.7cm][r]{\color{gray}{1.001}}\color{gray}{\times}\\{\mathmakebox[.7cm][r]{\scriptscriptstyle \color{gray}{\pm 0.0027}}}}$&&\cellcolor{black!5}{\tiny \color{gray}{p$_{28}$}:}&\cellcolor{black!5}$\substack{\mathmakebox[.7cm][r]{\color{gray}{0.247}}\color{gray}{s}\\{\mathmakebox[.7cm][r]{\scriptscriptstyle \color{gray}{\pm 0.0005}}}}$&\cellcolor{black!5}$\substack{\mathmakebox[.7cm][r]{\color{gray}{1.004}}\color{gray}{\times}\\{\mathmakebox[.7cm][r]{\scriptscriptstyle \color{gray}{\pm 0.0029}}}}$&&\cellcolor{black!5}{\tiny p$_{44}$:}&\cellcolor{black!5}$\substack{\mathmakebox[.7cm][r]{0.248}s\\{\mathmakebox[.7cm][r]{\scriptscriptstyle \pm 0.0005}}}$&\cellcolor{black!5}$\substack{\mathmakebox[.7cm][r]{1.011}\times\\{\mathmakebox[.7cm][r]{\scriptscriptstyle \pm 0.0031}}}$&&\cellcolor{black!5}{\tiny \color{gray}{p$_{60}$}:}&\cellcolor{black!5}$\substack{\mathmakebox[.7cm][r]{\color{gray}{0.247}}\color{gray}{s}\\{\mathmakebox[.7cm][r]{\scriptscriptstyle \color{gray}{\pm 0.0005}}}}$&\cellcolor{black!5}$\substack{\mathmakebox[.7cm][r]{\color{gray}{1.005}}\color{gray}{\times}\\{\mathmakebox[.7cm][r]{\scriptscriptstyle \color{gray}{\pm 0.0030}}}}$&&\cellcolor{black!5}{\tiny \color{gray}{p$_{76}$}:}&\cellcolor{black!5}$\substack{\mathmakebox[.7cm][r]{\color{gray}{0.246}}\color{gray}{s}\\{\mathmakebox[.7cm][r]{\scriptscriptstyle \color{gray}{\pm 0.0004}}}}$&\cellcolor{black!5}$\substack{\mathmakebox[.7cm][r]{\color{gray}{1.002}}\color{gray}{\times}\\{\mathmakebox[.7cm][r]{\scriptscriptstyle \color{gray}{\pm 0.0025}}}}$\\
{\tiny \color{gray}{p$_{13}$}:}&$\substack{\mathmakebox[.7cm][r]{\color{gray}{0.246}}\color{gray}{s}\\{\mathmakebox[.7cm][r]{\scriptscriptstyle \color{gray}{\pm 0.0005}}}}$&$\substack{\mathmakebox[.7cm][r]{\color{gray}{1.000}}\color{gray}{\times}\\{\mathmakebox[.7cm][r]{\scriptscriptstyle \color{gray}{\pm 0.0029}}}}$&&{\tiny \color{gray}{p$_{29}$}:}&$\substack{\mathmakebox[.7cm][r]{\color{gray}{0.246}}\color{gray}{s}\\{\mathmakebox[.7cm][r]{\scriptscriptstyle \color{gray}{\pm 0.0006}}}}$&$\substack{\mathmakebox[.7cm][r]{\color{gray}{1.000}}\color{gray}{\times}\\{\mathmakebox[.7cm][r]{\scriptscriptstyle \color{gray}{\pm 0.0032}}}}$&&{\tiny p$_{45}$:}&$\substack{\mathmakebox[.7cm][r]{0.248}s\\{\mathmakebox[.7cm][r]{\scriptscriptstyle \pm 0.0005}}}$&$\substack{\mathmakebox[.7cm][r]{1.011}\times\\{\mathmakebox[.7cm][r]{\scriptscriptstyle \pm 0.0027}}}$&&{\tiny p$_{61}$:}&$\substack{\mathmakebox[.7cm][r]{0.248}s\\{\mathmakebox[.7cm][r]{\scriptscriptstyle \pm 0.0005}}}$&$\substack{\mathmakebox[.7cm][r]{1.009}\times\\{\mathmakebox[.7cm][r]{\scriptscriptstyle \pm 0.0028}}}$&&{\tiny p$_{77}$:}&$\substack{\mathmakebox[.7cm][r]{0.244}s\\{\mathmakebox[.7cm][r]{\scriptscriptstyle \pm 0.0005}}}$&$\substack{\mathmakebox[.7cm][r]{0.992}\times\\{\mathmakebox[.7cm][r]{\scriptscriptstyle \pm 0.0027}}}$\\
\cellcolor{black!5}{\tiny p$_{14}$:}&\cellcolor{black!5}$\substack{\mathmakebox[.7cm][r]{0.248}s\\{\mathmakebox[.7cm][r]{\scriptscriptstyle \pm 0.0005}}}$&\cellcolor{black!5}$\substack{\mathmakebox[.7cm][r]{1.009}\times\\{\mathmakebox[.7cm][r]{\scriptscriptstyle \pm 0.0029}}}$&&\cellcolor{black!5}{\tiny \color{gray}{p$_{30}$}:}&\cellcolor{black!5}$\substack{\mathmakebox[.7cm][r]{\color{gray}{0.246}}\color{gray}{s}\\{\mathmakebox[.7cm][r]{\scriptscriptstyle \color{gray}{\pm 0.0006}}}}$&\cellcolor{black!5}$\substack{\mathmakebox[.7cm][r]{\color{gray}{1.000}}\color{gray}{\times}\\{\mathmakebox[.7cm][r]{\scriptscriptstyle \color{gray}{\pm 0.0031}}}}$&&\cellcolor{black!5}{\tiny p$_{46}$:}&\cellcolor{black!5}$\substack{\mathmakebox[.7cm][r]{0.270}s\\{\mathmakebox[.7cm][r]{\scriptscriptstyle \pm 0.0005}}}$&\cellcolor{black!5}$\substack{\mathmakebox[.7cm][r]{1.100}\times\\{\mathmakebox[.7cm][r]{\scriptscriptstyle \pm 0.0029}}}$&&\cellcolor{black!5}{\tiny \color{gray}{p$_{62}$}:}&\cellcolor{black!5}$\substack{\mathmakebox[.7cm][r]{\color{gray}{0.246}}\color{gray}{s}\\{\mathmakebox[.7cm][r]{\scriptscriptstyle \color{gray}{\pm 0.0005}}}}$&\cellcolor{black!5}$\substack{\mathmakebox[.7cm][r]{\color{gray}{1.000}}\color{gray}{\times}\\{\mathmakebox[.7cm][r]{\scriptscriptstyle \color{gray}{\pm 0.0027}}}}$&&\cellcolor{black!5}{\tiny \color{gray}{p$_{78}$}:}&\cellcolor{black!5}$\substack{\mathmakebox[.7cm][r]{\color{gray}{0.246}}\color{gray}{s}\\{\mathmakebox[.7cm][r]{\scriptscriptstyle \color{gray}{\pm 0.0003}}}}$&\cellcolor{black!5}$\substack{\mathmakebox[.7cm][r]{\color{gray}{0.999}}\color{gray}{\times}\\{\mathmakebox[.7cm][r]{\scriptscriptstyle \color{gray}{\pm 0.0025}}}}$\\
{\tiny \color{gray}{p$_{15}$}:}&$\substack{\mathmakebox[.7cm][r]{\color{gray}{0.246}}\color{gray}{s}\\{\mathmakebox[.7cm][r]{\scriptscriptstyle \color{gray}{\pm 0.0004}}}}$&$\substack{\mathmakebox[.7cm][r]{\color{gray}{0.999}}\color{gray}{\times}\\{\mathmakebox[.7cm][r]{\scriptscriptstyle \color{gray}{\pm 0.0026}}}}$&&{\tiny \color{gray}{p$_{31}$}:}&$\substack{\mathmakebox[.7cm][r]{\color{gray}{0.246}}\color{gray}{s}\\{\mathmakebox[.7cm][r]{\scriptscriptstyle \color{gray}{\pm 0.0004}}}}$&$\substack{\mathmakebox[.7cm][r]{\color{gray}{1.000}}\color{gray}{\times}\\{\mathmakebox[.7cm][r]{\scriptscriptstyle \color{gray}{\pm 0.0027}}}}$&&{\tiny p$_{47}$:}&$\substack{\mathmakebox[.7cm][r]{0.267}s\\{\mathmakebox[.7cm][r]{\scriptscriptstyle \pm 0.0004}}}$&$\substack{\mathmakebox[.7cm][r]{1.085}\times\\{\mathmakebox[.7cm][r]{\scriptscriptstyle \pm 0.0027}}}$&&{\tiny p$_{63}$:}&$\substack{\mathmakebox[.7cm][r]{0.245}s\\{\mathmakebox[.7cm][r]{\scriptscriptstyle \pm 0.0004}}}$&$\substack{\mathmakebox[.7cm][r]{0.999}\times\\{\mathmakebox[.7cm][r]{\scriptscriptstyle \pm 0.0026}}}$&&{\tiny \color{gray}{p$_{79}$}:}&$\substack{\mathmakebox[.7cm][r]{\color{gray}{0.246}}\color{gray}{s}\\{\mathmakebox[.7cm][r]{\scriptscriptstyle \color{gray}{\pm 0.0004}}}}$&$\substack{\mathmakebox[.7cm][r]{\color{gray}{0.999}}\color{gray}{\times}\\{\mathmakebox[.7cm][r]{\scriptscriptstyle \color{gray}{\pm 0.0026}}}}$\\
\cellcolor{black!5}{\tiny \color{gray}{p$_{16}$}:}&\cellcolor{black!5}$\substack{\mathmakebox[.7cm][r]{\color{gray}{0.246}}\color{gray}{s}\\{\mathmakebox[.7cm][r]{\scriptscriptstyle \color{gray}{\pm 0.0004}}}}$&\cellcolor{black!5}$\substack{\mathmakebox[.7cm][r]{\color{gray}{1.000}}\color{gray}{\times}\\{\mathmakebox[.7cm][r]{\scriptscriptstyle \color{gray}{\pm 0.0026}}}}$&&\cellcolor{black!5}{\tiny \color{gray}{p$_{32}$}:}&\cellcolor{black!5}$\substack{\mathmakebox[.7cm][r]{\color{gray}{0.246}}\color{gray}{s}\\{\mathmakebox[.7cm][r]{\scriptscriptstyle \color{gray}{\pm 0.0005}}}}$&\cellcolor{black!5}$\substack{\mathmakebox[.7cm][r]{\color{gray}{1.000}}\color{gray}{\times}\\{\mathmakebox[.7cm][r]{\scriptscriptstyle \color{gray}{\pm 0.0028}}}}$&&\cellcolor{black!5}{\tiny \color{gray}{p$_{48}$}:}&\cellcolor{black!5}$\substack{\mathmakebox[.7cm][r]{\color{gray}{0.247}}\color{gray}{s}\\{\mathmakebox[.7cm][r]{\scriptscriptstyle \color{gray}{\pm 0.0005}}}}$&\cellcolor{black!5}$\substack{\mathmakebox[.7cm][r]{\color{gray}{1.003}}\color{gray}{\times}\\{\mathmakebox[.7cm][r]{\scriptscriptstyle \color{gray}{\pm 0.0030}}}}$&&\cellcolor{black!5}{\tiny \color{gray}{p$_{64}$}:}&\cellcolor{black!5}$\substack{\mathmakebox[.7cm][r]{\color{gray}{0.245}}\color{gray}{s}\\{\mathmakebox[.7cm][r]{\scriptscriptstyle \color{gray}{\pm 0.0005}}}}$&\cellcolor{black!5}$\substack{\mathmakebox[.7cm][r]{\color{gray}{0.998}}\color{gray}{\times}\\{\mathmakebox[.7cm][r]{\scriptscriptstyle \color{gray}{\pm 0.0028}}}}$&\\
\bottomrule
\end{tabular}
\end{adjustbox}
\vspace{0.5em}

\caption{DeltaBlue permutations in \ourvm, with absolute times (in seconds) and
relative timings (to mono-PHP DeltaBlue run on \ourvm).
Greyed-out cells indicate that the confidence intervals overlap. Bold
entries indicate that there is more than a 25\% relative performance
difference.}
\label{tab:db_perms_results}
\end{table*}
\addtolength{\tabcolsep}{.4em}
Table~\ref{tab:db_perms_results} shows the results from the permutations
experiment. The majority of permutations are statistically indistinguishable
from mono-PHP; most of the remainder are close enough
in performance to be of little interest. Four permutations, however, show substantial
differences: $p_2, p_5, p_6$ and $p_7$ all perform much better than in mono-PHP.
We now describe the reasons for these cases.

$p_2$ swaps the \texttt{OrderedCollection} class's constructor which performs a
single action, assigning an array (in PHP) or list (in Python) to the
\texttt{elms} attribute. The seemingly innocuous change of moving from a PHP
array to a Python list has a big impact on performance simply because \pypy's
lists are are far more extensively optimised than \hippy's
(see~\cite{bolz13strategies}). This provides indirect evidence of the importance
of making adapters immutable (see Section~\ref{sec:designconversions}): even
though $p_2$ operates extensively on adapters, their costs
after trace optimisation are extremely small.

$p_5$, $p_6$, and $p_7$ are all similar in nature. Ultimately, and perhaps
surprisingly, the slowdown is due to Hippy using a tracing garbage collector.
Because of PHP's copy-on-write semantics, arrays are
conceptually copied on every mutation. Zend (the traditional PHP implementation)
is able to optimise away many of these writes by making use of its reference
counting garbage collector. When a mutation operation occurs on an array with a reference count of 1,
the array is mutated in place, as the change cannot be
observed elsewhere. \hippy in contrast does not use reference counting, and does
not know exactly how many pointers to an array exist at any given point.
Checking every pointer in the run-time system would be prohibitively expensive,
so \hippy approximates Zend's optimisation with a `unique' flag on
array references. Various operations can remove uniqueness, but arrays
in unique references can be optimised in the same manner as Zend arrays with a
reference count of 1. Taking $p_5$ as a concrete
example, we can see the subtle effects of this optimisation in \hippy. $p_5$ swaps
the \texttt{OrderedCollection} class's \texttt{size} method, which simply calls
\texttt{count} (in PHP) or \texttt{len} (in Python) on a PHP array stored in an
attribute. Since the \texttt{count} function is call-by-value, \hippy optimises
the copy that of the array that should occur by simply dropping its unique flag;
later mutations thus must therefore copy the array.
However, this is not directly why \ourvm is faster than \hippy in $p_5$.
When the \texttt{OrderedCollection} class's \texttt{size} method is moved to
Python, \ourvm's mutability semantics (see Section~\ref{sec:arrayrefs})
cause \texttt{size} to be a pass-by-reference function, thus meaning that
the PHP reference does not lose uniqueness, and less copying is then required.

\subsection{Threats to Validity}

Benchmarks are only ever a snapshot of certain performance characteristics of a
system, and we do not pretend that they necessarily tell us about program
performance in a more general setting. Our experiments also make no attempt to
account for JIT warmup (for reasons explained in Section~\ref{no warmup
chopping}). Removing JIT warmup would thus `improve' the perceived timings of
VMs such as \ourvm which perform JIT compilation. Since it is also known that
RPython VMs have relatively poor warmup~\cite{bolz14impact}, the likely effect of
our decision is to make \ourvm look worse relative to other VMs. We consider
this a better trade-off than trying to make other VMs look worse relative to
\ourvm.

\section{Case Studies}
\label{sec:case studies}

\begin{figure*}[t]
\centering
\includegraphics[width=1.2\textwidth, center]{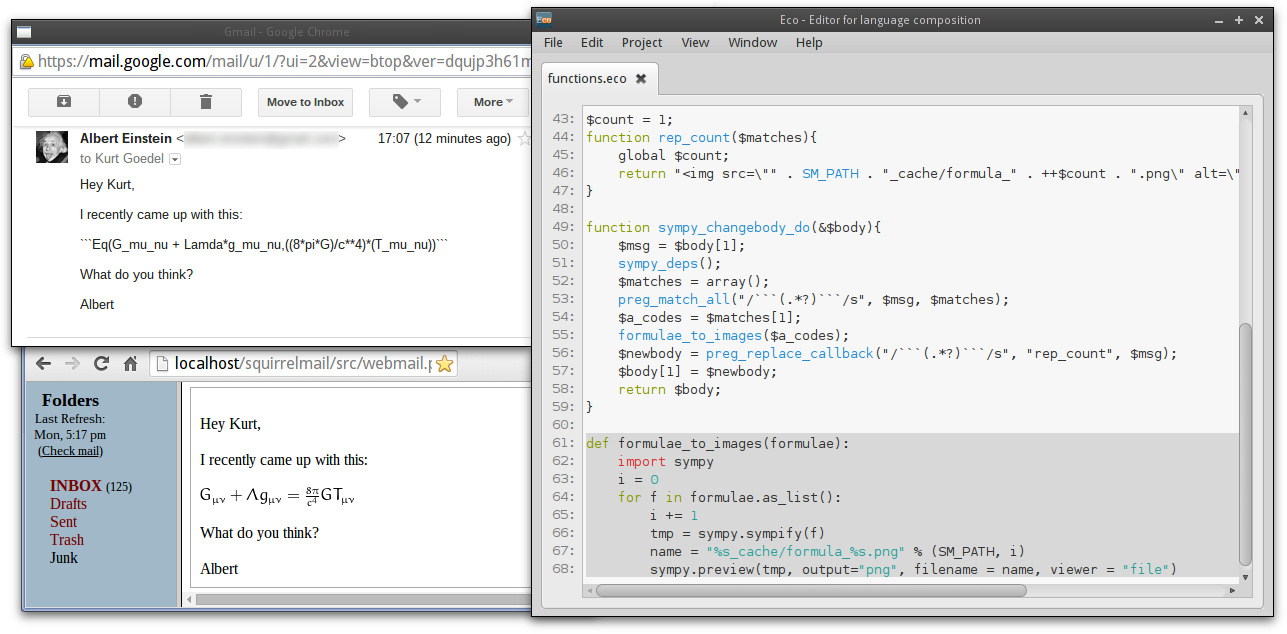}
\caption{Example mails sent with our extended version of SquirrelMail. We
extended this PHP mail client such that it can visualise mathematical formulae
using the SymPy Python library. A portion of the plug-in code is shown in the
right.}

\label{fig:sympyexample}
\end{figure*}

\subsection{Using CFFI in PHP}

PHP does not have a built-in C FFI, whereas Python does via the \texttt{cffi}
module. PHP code can thus use \ourvm to access \texttt{cffi},
acquiring a C FFI by default. For example the following elided example
shows PHP using \texttt{cffi} to call the Unix \texttt{clock\_gettime}
function:
\begin{lstlisting}[numbers=left]
$cffi = import_py_mod("cffi");
$ffi = new $cffi->FFI();
$ffi->cdef("double _clock_gettime_monotonic();");
$csrc = <<<EOD
  double _clock_gettime_monotonic(){
    struct timespec ts;
    if ((clock_gettime(CLOCK_MONOTONIC, &ts)) == -1)
      err(1, "clock_gettime error");
    return ts.tv_sec + ts.tv_nsec * pow(10, -9);
  }
EOD;
$ffi->set_source("_example", $csrc);
$C = $ffi->dlopen(null);
echo "Monotonic time: " . $C->_clock_gettime_monotonic() . "\n";
\end{lstlisting}

\subsection{A SquirrelMail Plugin}

SquirrelMail is a venerable PHP web mail client. We used
\ourvm to add a SquirrelMail plug-in that uses the Python SymPy library.
This is intended to show
that \ourvm can be used to add Python modules to relatively large existing systems. In
essence, the plug-in recognises mathematical formulae between triple backticks,
and uses SymPy to render them in traditional mathematical notation. Formulae
in incoming emails are automatically rendered; users
sending emails with such formulae can preview the rendering before sending.
Figure~\ref{fig:sympyexample} shows the plug-in in use, and the core parts of the
code within Eco.

The \texttt{sym\-py\-\_changebody\_do} function is called by SquirrelMail's
\texttt{mess\-age\_body} hook (which is also called upon viewing a message),
receiving the content of the email as an argument. A regular expression finds
all occurrences of formulae between backticks (line 53) and passes them to the
Python \texttt{formulae\-\_to\-\_images} function. This then uses SymPy to convert
the formulae to images (numbered by their offset in the array/list) into the
directory pointed to by the PHP constant \texttt{SM\_PATH} (lines 61--68), and
uses the URL of the image in-place of the textual formula (line 56).

\subsection{System Language Migration}
\label{sec:migration}

We expect that one of the key uses of syntactic language composition
is system language migration, where systems are slowly migrated from language
$A$ to $B$ in small stages. Instead of having to rewrite whole modules or
sub-systems, syntactic language composition offers the possibility of migrating
one function at a time. A full case study is far beyond the scope of this
paper, but we implicitly modelled this technique when creating the DeltaBlue and
Richards benchmarks, where we translated each PHP method into Python, leaving only
`shell' PHP classes, global functions and variables. As
Section~\ref{sec:results} clearly shows, the resulting performance is at worst
2x of its mono-language variant, which we believe makes system language migration plausible
for the first time.

\section{Discussion}

To give an approximate idea of \ourvm's size, some rough metrics are useful. The
\texttt{pypy\_\-bridge} module -- in which the majority of \ourvm is implemented
-- adds around 2KLoC. Aside from this we added around
0.25KLoC and 0.2KLoC to the existing \hippy and \pypy interpreter code respectively.
5KLoC of new unit tests were added. On a fast build machine (4GHz Core i7)
\ourvm takes about 45 minutes to build. We estimate that implementing \ourvm took
around 7 person months.

A succinct summary of our experiences of creating \ourvm is: implementing what
we wanted was fairly easy; making what we implemented run fast was somewhat easy;
but working out what we wanted to implement was often hard. The latter point may
surprise readers as much as it has surprised us. There are two main reasons for
this.

First, there is little precedent for fine-grained syntactic language
composition. Most existing language compositions are either extremely crude (per
process compositions) or have design decisions implicitly imposed upon them
(translating into another VM's bytecode). We therefore faced a number of novel
language design issues, and used gradually larger case studies to help us
iterate our way to good solutions, sometimes exhausting what felt like
every possible alternative. Cross-language scoping is a good example of this: we
tried many possibilities before settling on the scheme described in
Section~\ref{sec:xscope}.

Second, it is difficult, and probably impossible,
for any single person to be truly expert in every language and implementation involved in a
composition. We sometimes had to base initial designs on (hopefully) intelligent guesses about
one or the other languages' semantics or implementations. We
then tried as hard as we could to break the resulting design. It rapidly became
clear to us that in large languages such as PHP and Python, there are many
corner-cases, sometimes little used, which need to be considered. PHP references
caused us more headaches than any other language feature. At first we ignored
them, and then we failed to appreciate their pervasive nature.
It took us considerable effort to understand them well enough to
make sense of their place within \ourvm. While we do
not pretend to be experts in every aspect of either PHP or Python, we can recommend this route
to anyone who wishes to understand the nooks and crannies of a language and its
implementation.

Once we had settled upon a good design, we rarely had substantial difficulty in
modifying \hippy or \pypy to implement it. The relatively small size of the
additional / changed code in the composition is a reasonable proxy for this.
Similarly, the very nature of
meta-tracing meant that most cross-language optimisations came without any extra
work on our part. Cross-language variable scoping was the only feature that
required substantial optimisation effort on our part, including to \hippy
itself.

\subsection{Generalising from the Case Study}

\ourvm is, to the best of our knowledge, the first fine-grained language
composition. Although we are cautious about over-generalising our results, we
believe that some of the lessons embodied in \ourvm may be relevant for future
fine-grained language compositions.

Most obviously, despite the rather different run-time properties of PHP and
Python, \ourvm's performance is close enough to \hippy and \pypy to be usable.
While we would like to claim credit for all of this, most of the benefit comes
from meta-tracing: only in a few places did we have to add \ourvm-specific
optimisations. We expect languages even more disparate than PHP and Python
to still achieve fairly good performance using meta-tracing.

Our use of adapters meant that most interactions between PHP and Python required
little or no effort on our part to compose together satisfactorily. We expect
this to generalise to most other compositions. Adapters were also the key to
resolving seemingly major semantic data-type incompatibilities between (mostly
immutable) PHP and (mostly mutable) Python. The techniques we used are likely to
be relevant to compositions involving languages that are more rigorously
immutable than PHP.

Finally, despite the archaic nature of PHP and Python's scoping rules, we were
able to design good cross-language scoping rules. Most modern languages have
embraced lexical scoping, and compositions involving them will require less
contortions than \ourvm.

\section{Related Work}
\label{sec:related}

There has been a long-standing desire for language composition (see
e.g.~\cite{cheatham69motivation}), and many flavours have developed since then.
Extensible languages
(e.g.~\cite{irons70experience,gregory85ametalanguage,cardelli93extensible}) aim
to grow a language as required by a user. However, the base language
places restrictions on what extensions are possible (e.g.~due to parsing
restrictions) and performant~\cite{tratt08domainspecific}. Translating one
language into another (with e.g.~Stratego~\cite{bravenboer08stratego}) removes
many of the limitations on what is expressible, but  full-scale translations are
complex (e.g.~\cite{gray05finegrained}) and typically suffer the same
performance issues as extensible languages. However, for small use cases, or
where performance is not important, either approach can work well.

FFIs are the most common approach to composing languages, but their performance
is typically poor due to their inability to inline cross-language calls. The
next most common mechanism is to target an existing high performance VM
(typically HotSpot). However, since such VMs can only optimise those programs
they expect to commonly see. Languages which step even slightly outside this
mould perform poorly. For example, Java programs have often excellent
performance on HotSpot, but Python programs on HotSpot generally run slower than
with simple C-based interpreters~\cite{sarimbekov13characteriustics,bolz14impact}.

Our aim in this paper has been to show that fine-grained syntactic language
composition is possible and performant. We make no claims about the formal
properties of the resulting composition as the practical challenges identified
in this paper are already substantial. There is already a small body of work on
formalising language composition, such as an investigation of the COM
architecture~\cite{sullivan99analysis}, and an abstract framework for
specifying the operational semantics of multi-language
embeddings~\cite{matthews09operational}. There are also partial formal semantics
for languages such as Python
and PHP (see e.g.~\cite{politz13monty,filaretti14phpsemantics}). We welcome
future work formalising fine-grained compositions.

As the case studies show, \ourvm is at least somewhat usable, but we are under
no illusions that it is an industrial strength product. There are many interesting
directions for further exploration, such as experimenting with cross-language
inheritance~\cite{gray08safe}.

\section{Conclusions}
\label{sec:conclusion}

In this paper we introduced \ourvm, a fine-grained syntactic composition of PHP
and Python implemented by combining together meta-tracing interpreters. We
consider that \ourvm validates our hypothesis that programming languages can be
composed in a finer-grained manner than previously thought possible or practical. Not only does \ourvm
introduce novel concepts such as cross-language variable scoping, but its
performance is close enough to its mono-language cousins to encourage use of
such a system. Inevitably, some of \ourvm's details are specific to the
particular pair of languages it composes. However, many of the techniques that
\ourvm embodies -- the use of interpreter composition with meta-tracing; some of
the design choices surrounding cross-language scoping -- are likely to be of use
to future language compositions.

\subparagraph*{Acknowledgements}

We thank Armin Rigo for
adjusting RPython to cope with some of \ourvm's demands, and advice on Hippy;
Ronan Lamy and Maciej Fijałkowski for help with Hippy; Jasper Schulz for
help with cross-language exceptions; Alan Mycroft
for insightful thoughts on language composition; and Martin Berger, Darya
Kurilova, and Sarah Mount for comments. This research
was funded by the EPSRC COOLER (EP/K01790X/1) and LECTURE (EP/L02344X/1) grants.

\bibliographystyle{plain}
\bibliography{barrett}

\begin{thebibliography}{10}

\bibitem{anderson94fannkuch}
Kenneth~R. Anderson and Duane Rettig.
\newblock Performing {Lisp} analysis of the {Fannkuch} benchmark.
\newblock {\em Lisp Pointers}, 7(4):2--12, Oct 1994.

\bibitem{bala00dynamo}
Vasanth Bala, Evelyn Duesterwald, and Sanjeev Banerjia.
\newblock Dynamo: A transparent dynamic optimization system.
\newblock In {\em PLDI}, pages 1--12, Jun 2000.

\bibitem{barrett15approaches}
Edd Barrett, Carl~Friedrich Bolz, and Laurence Tratt.
\newblock Approaches to interpreter composition.
\newblock {\em COMLAN}, 44(C), March 2015.

\bibitem{bebenita10spur}
Michael Bebenita, Florian Brandner, Manuel Fahndrich, Francesco Logozzo,
  Wolfram Schulte, Nikolai Tillmann, and Herman Venter.
\newblock {SPUR}: A trace-based {JIT} compiler for {CIL}.
\newblock In {\em OOPSLA}, pages 708--725, Mar 2010.

\bibitem{bolz13strategies}
Carl~Friedrich Bolz, Lukas Diekmann, and Laurence Tratt.
\newblock Storage strategies for collections in dynamically typed languages.
\newblock In {\em OOPSLA}, pages 167--182, Oct 2013.

\bibitem{bolz14impact}
Carl~Friedrich Bolz and Laurence Tratt.
\newblock The impact of meta-tracing on {VM} design and implementation.
\newblock {\em SCICO}, 98(3):408--421, Feb 2015.

\bibitem{bravenboer08stratego}
Martin Bravenboer, Karl~Trygve Kalleberg, Rob Vermaas, and Eelco Visser.
\newblock Stratego/{XT} 0.17. {A} language and toolset for program
  transformation.
\newblock {\em SCICO}, 72(1–2):52 -- 70, 2008.

\bibitem{cardelli93extensible}
Luca Cardelli, Florian Matthes, and Mart\'{\i}n Abadi.
\newblock Extensible grammars for language specialization.
\newblock In {\em Database Programming Languages}, pages 11--31, Aug 1993.

\bibitem{chambers89efficient}
Craig Chambers, David Ungar, and Elgin Lee.
\newblock An efficient implementation of {SELF} a dynamically-typed
  object-oriented language based on prototypes.
\newblock In {\em OOPSLA}, Sep 1989.

\bibitem{cheatham69motivation}
Thomas~E. Cheatham.
\newblock Motivation for extensible languages.
\newblock {\em SIGPLAN}, 4(8):45--49, Aug 1969.

\bibitem{diekmann14eco}
Lukas Diekmann and Laurence Tratt.
\newblock {Eco}: A language composition editor.
\newblock In {\em SLE}, pages 82--101, Sep 2014.

\bibitem{filaretti14phpsemantics}
Daniele Filaretti and Sergio Maffeis.
\newblock An executable formal semantics of {PHP}.
\newblock In {\em ECOOP}, pages 567--592, 2014.

\bibitem{gal06hotpathvm}
Andreas Gal, Christian~W. Probst, and Michael Franz.
\newblock {HotpathVM:} an effective {JIT} compiler for resource-constrained
  devices.
\newblock In {\em VEE}, pages 144--153, Jun 2006.

\bibitem{gray08safe}
Kathryn~E. Gray.
\newblock Safe cross-language inheritance.
\newblock In {\em ECOOP}, pages 52--75, jul 2008.

\bibitem{gray05finegrained}
Kathryn~E. Gray, Robert~Bruce Findler, and Matthew Flatt.
\newblock Fine-grained interoperability through mirrors and contracts.
\newblock In {\em OOPSLA}, pages 231--245, Oct 2005.

\bibitem{grimmer2014dynamically}
Mathias Grimmer, Chris Seaton, Thomas W\"urthinger, and Hanspeter
  M\"ossenb\"ock.
\newblock Dynamically composing languages in a modular way: Supporting {C}
  extensions for dynamic languages.
\newblock In {\em Modularity}, March 2015.

\bibitem{grimmer15interop}
Matthias Grimmer, Chris Seaton, Roland Schatz, Thomas W\"{u}rthinger, and
  Hanspeter M\"{o}ssenb\"{o}ck.
\newblock High-performance cross-language interoperability in a multi-language
  runtime.
\newblock In {\em DLS}, pages 78--90, 2015.

\bibitem{ingalls97back}
Dan Ingalls, Ted Kaehler, John Maloney, Scott Wallace, and Alan Kay.
\newblock Back to the future: the story of {Squeak}, a practical {Smalltalk}
  written in itself.
\newblock In {\em OOPSLA}, pages 318--326, Oct 1997.

\bibitem{irons70experience}
Edgar~T. Irons.
\newblock Experience with an extensible language.
\newblock {\em CACM}, 13(1):31--40, Jan 1970.

\bibitem{gregory85ametalanguage}
Gregory~F. Johnson and C.~N. Fischer.
\newblock A meta-language and system for nonlocal incremental attribute
  evaluation in language-based editors.
\newblock In {\em POPL}, pages 141--151, Jan 1985.

\bibitem{kalibera12quantifying}
Tomas Kalibera and Richard Jones.
\newblock Quantifying performance changes with effect size confidence
  intervals.
\newblock Technical Report 4-12, University of Kent, Jun 2012.

\bibitem{kelsey94tractable}
Richard~A. Kelsey and Jonathan~A. Rees.
\newblock A tractable {Scheme} implementation.
\newblock {\em Lisp Symb. Comput.}, 7(4):315--335, Dec 1994.

\bibitem{matthews09operational}
Jacob Matthews and Robert~Bruce Findler.
\newblock Operational semantics for multi-language programs.
\newblock {\em TOPLAS}, 31(3):12:1--12:44, Apr 2009.

\bibitem{mitchell70design}
James~George Mitchell.
\newblock {\em The design and construction of flexible and efficient
  interactive programming systems}.
\newblock PhD thesis, Carnegie Mellon University, Jun 1970.

\bibitem{oneill15pcg}
Melissa~E. O'Neil.
\newblock {PCG}: A family of simple fast space-efficient statistically good
  algorithms for random number generation, 2015.

\bibitem{politz13monty}
Joe~Gibbs Politz, Alejandro Martinez, Matthew Milano, Sumner Warren, Daniel
  Patterson, Junsong Li, Anand Chitipothu, and Shriram Krishnamurthi.
\newblock Python: The full {Monty}.
\newblock In {\em OOPSLA}, pages 217--232, 2013.

\bibitem{sannella93deltablue}
Michael Sannella, John Maloney, Bjorn Freeman-Benson, and Alan Borning.
\newblock Multi-way versus one-way constraints in user interfaces: Experience
  with the {DeltaBlue} algorithm.
\newblock {\em SPE}, 23(5):529--566, 1993.

\bibitem{sarimbekov13characteriustics}
Aibek Sarimbekov, Andrej Podzimek, Lubomir Bulej, Yudi Zheng, Nathan Ricci, and
  Walter Binder.
\newblock Characteristics of dynamic {JVM} languages.
\newblock In {\em VMIL}, pages 11--20, Oct 2013.

\bibitem{sullivan03dynamic}
Gregory~T. Sullivan, Derek~L. Bruening, Iris Baron, Timothy Garnett, and Saman
  Amarasinghe.
\newblock Dynamic native optimization of interpreters.
\newblock In {\em IVME}, pages 50--57, Jun 2003.

\bibitem{sullivan99analysis}
Kevin~J. Sullivan, Mark Marchukov, and John Socha.
\newblock Analysis of a conflict between aggregation and interface negotiation
  in {Microsoft}'s component object model.
\newblock {\em TOSE}, 25(4):584--599, Jul 1999.

\bibitem{tratt08domainspecific}
Laurence Tratt.
\newblock Domain specific language implementation via compile-time
  meta-programming.
\newblock {\em TOPLAS}, 30(6):1--40, Oct 2008.

\bibitem{wuerthinger13onevm}
Thomas W\"{u}rthinger, Christian Wimmer, Andreas W\"{o}\ss, Lukas Stadler,
  Gilles Duboscq, Christian Humer, Gregor Richards, Doug Simon, and Mario
  Wolczko.
\newblock One {VM} to rule them all.
\newblock In {\em Onward!}, pages 187--204, 2013.

\bibitem{yermolovich09optimization}
Alexander Yermolovich, Christian Wimmer, and Michael Franz.
\newblock Optimization of dynamic languages using hierarchical layering of
  virtual machines.
\newblock In {\em DLS}, pages 79--88, Oct 2009.

\end{thebibliography}

\newpage

\appendix

\section{Small Microbenchmarks}
\label{sec:microbenchmarks}

Section~\ref{microbenchmarks} outlined the small microbenchmarks used in our experiment.
This appendix lists and describes each of the small microbenchmarks. The following
microbenchmarks consist of an outer loop calling an inner function:
\begin{description*}
\item[l1a0r] The inner function takes an integer which is decremented to
    zero in a loop. Nothing is returned.
\item[l1a1r] The inner function takes an integer which is decremented to
zero in a loop. After every decrement, the value is added to a sum total. The sum is returned.
\item[ref\_swap] The inner function swaps its two
    arguments using references. Since Python has no
    support for references, there is no mono-Python variant of this benchmark.
\item[return\_simple] The inner function returns a constant integer.
\item[scopes] The inner function takes a parameter and adds it to a variable
    from an outer scope. In the composed variants,
    the scope lookup crosses language boxes.
\item[smallfunc] The inner function takes three arguments $a$, $b$, and
    $c$, and returns $a + b * c$.
\item[sum] The inner function takes five arguments, sums them, and returns the result.
\item[sum\_meth] As \emph{sum}, except the sum is computed and returned by a method. The method belongs
to an object which is allocated once and re-used.
\item[sum\_meth\_attr] As \emph{sum\_meth}, except that the result is stored to an attribute of the object.
\item[total\_list] The inner function sums the elements of a list/array passed as an argument.
\end{description*}
The following microbenchmarks consist of one function generating
elements which another function consumes:
\begin{description*}
\item[instchain] A nested chain of objects is constructed and consumed in a
loop. Each object in the chain has an attribute storing an integer. One
function constructs the chain, another walks it summing the integers. These
functions are called in a loop. In the
composed variants, the outer loop is in one language and the construct and
walk functions (including utility methods) are in the other.
\item[lists] One function constructs a list of integers, and another iterates
over the list, summing its elements. These functions are called in a loop to repeatedly
construct and sum lists. In the composed variant, the summing function is
written in one language, with all other parts written in the other.
\item[list\_walk] One function creates a linked list while the other function
walks the list. Each element in the list is a three element tuple $(x, y, n)$ where
$x$ and $y$ are integers and $n$ is a pointer to the next element, or
the string \texttt{"end"} for the final element. As the list is walked, a counter is
incremented by $y-x$.  In the composed variant, the list creation and walking
functions are in a different language from the outer loop.
\end{description*}
The l1a0r, l1a1r, lists, smallfunc, and list\_walk microbenchmarks are
ports of benchmarks from~\cite{barrett15approaches}. All other small microbenchmarks
were created
specifically to test \ourvm.

\begin{table*}
\centering

\begin{adjustbox}{width=1.2\textwidth,center}
\begin{tabular}{lrrrrrrrr}
\toprule
Benchmark&\makebox[1.45cm][r]{CPython}&\makebox[1.45cm][r]{HHVM}&\makebox[1.45cm][r]{HippyVM}&\makebox[1.45cm][r]{PyHyp$_\textrm{PHP}$}&\makebox[1.45cm][r]{PyHyp$_\textrm{Py}$}&\makebox[1.45cm][r]{PyHyp$_\textrm{mono}$}&\makebox[1.45cm][r]{PyPy}&\makebox[1.45cm][r]{Zend}\\
\toprule
instchain&$\substack{\mathmakebox[.7cm][r]{11.323}\\{\mathmakebox[.7cm][r]{\scriptscriptstyle \pm 0.0208}}}$&$\substack{\mathmakebox[.7cm][r]{3.232}\\{\mathmakebox[.7cm][r]{\scriptscriptstyle \pm 0.0016}}}$&$\substack{\mathmakebox[.7cm][r]{0.309}\\{\mathmakebox[.7cm][r]{\scriptscriptstyle \pm 0.0002}}}$&$\substack{\mathmakebox[.7cm][r]{0.338}\\{\mathmakebox[.7cm][r]{\scriptscriptstyle \pm 0.0003}}}$&&$\substack{\mathmakebox[.7cm][r]{0.378}\\{\mathmakebox[.7cm][r]{\scriptscriptstyle \pm 0.0002}}}$&$\substack{\mathmakebox[.7cm][r]{0.228}\\{\mathmakebox[.7cm][r]{\scriptscriptstyle \pm 0.0001}}}$&$\substack{\mathmakebox[.7cm][r]{12.345}\\{\mathmakebox[.7cm][r]{\scriptscriptstyle \pm 0.0519}}}$\\
\addlinespace
l1a0r&$\substack{\mathmakebox[.7cm][r]{15.965}\\{\mathmakebox[.7cm][r]{\scriptscriptstyle \pm 0.0011}}}$&$\substack{\mathmakebox[.7cm][r]{0.752}\\{\mathmakebox[.7cm][r]{\scriptscriptstyle \pm 0.0003}}}$&$\substack{\mathmakebox[.7cm][r]{0.254}\\{\mathmakebox[.7cm][r]{\scriptscriptstyle \pm 0.0000}}}$&$\substack{\mathmakebox[.7cm][r]{0.186}\\{\mathmakebox[.7cm][r]{\scriptscriptstyle \pm 0.0000}}}$&$\substack{\mathmakebox[.7cm][r]{0.252}\\{\mathmakebox[.7cm][r]{\scriptscriptstyle \pm 0.0000}}}$&$\substack{\mathmakebox[.7cm][r]{0.252}\\{\mathmakebox[.7cm][r]{\scriptscriptstyle \pm 0.0000}}}$&$\substack{\mathmakebox[.7cm][r]{0.249}\\{\mathmakebox[.7cm][r]{\scriptscriptstyle \pm 0.0019}}}$&$\substack{\mathmakebox[.7cm][r]{7.198}\\{\mathmakebox[.7cm][r]{\scriptscriptstyle \pm 0.0005}}}$\\
\addlinespace
l1a1r&$\substack{\mathmakebox[.7cm][r]{16.473}\\{\mathmakebox[.7cm][r]{\scriptscriptstyle \pm 0.0162}}}$&$\substack{\mathmakebox[.7cm][r]{0.586}\\{\mathmakebox[.7cm][r]{\scriptscriptstyle \pm 0.0001}}}$&$\substack{\mathmakebox[.7cm][r]{0.257}\\{\mathmakebox[.7cm][r]{\scriptscriptstyle \pm 0.0000}}}$&$\substack{\mathmakebox[.7cm][r]{0.197}\\{\mathmakebox[.7cm][r]{\scriptscriptstyle \pm 0.0002}}}$&$\substack{\mathmakebox[.7cm][r]{0.256}\\{\mathmakebox[.7cm][r]{\scriptscriptstyle \pm 0.0000}}}$&$\substack{\mathmakebox[.7cm][r]{0.256}\\{\mathmakebox[.7cm][r]{\scriptscriptstyle \pm 0.0000}}}$&$\substack{\mathmakebox[.7cm][r]{0.224}\\{\mathmakebox[.7cm][r]{\scriptscriptstyle \pm 0.0003}}}$&$\substack{\mathmakebox[.7cm][r]{7.689}\\{\mathmakebox[.7cm][r]{\scriptscriptstyle \pm 0.0166}}}$\\
\addlinespace
lists&$\substack{\mathmakebox[.7cm][r]{3.871}\\{\mathmakebox[.7cm][r]{\scriptscriptstyle \pm 0.0021}}}$&$\substack{\mathmakebox[.7cm][r]{0.448}\\{\mathmakebox[.7cm][r]{\scriptscriptstyle \pm 0.0015}}}$&$\substack{\mathmakebox[.7cm][r]{0.469}\\{\mathmakebox[.7cm][r]{\scriptscriptstyle \pm 0.0005}}}$&$\substack{\mathmakebox[.7cm][r]{0.481}\\{\mathmakebox[.7cm][r]{\scriptscriptstyle \pm 0.0008}}}$&$\substack{\mathmakebox[.7cm][r]{0.269}\\{\mathmakebox[.7cm][r]{\scriptscriptstyle \pm 0.0003}}}$&$\substack{\mathmakebox[.7cm][r]{0.470}\\{\mathmakebox[.7cm][r]{\scriptscriptstyle \pm 0.0005}}}$&$\substack{\mathmakebox[.7cm][r]{0.239}\\{\mathmakebox[.7cm][r]{\scriptscriptstyle \pm 0.0002}}}$&$\substack{\mathmakebox[.7cm][r]{7.036}\\{\mathmakebox[.7cm][r]{\scriptscriptstyle \pm 0.0136}}}$\\
\addlinespace
ref\_swap&&$\substack{\mathmakebox[.7cm][r]{2.573}\\{\mathmakebox[.7cm][r]{\scriptscriptstyle \pm 0.0001}}}$&$\substack{\mathmakebox[.7cm][r]{0.306}\\{\mathmakebox[.7cm][r]{\scriptscriptstyle \pm 0.0001}}}$&$\substack{\mathmakebox[.7cm][r]{0.306}\\{\mathmakebox[.7cm][r]{\scriptscriptstyle \pm 0.0000}}}$&$\substack{\mathmakebox[.7cm][r]{0.215}\\{\mathmakebox[.7cm][r]{\scriptscriptstyle \pm 0.0000}}}$&$\substack{\mathmakebox[.7cm][r]{0.306}\\{\mathmakebox[.7cm][r]{\scriptscriptstyle \pm 0.0000}}}$&&$\substack{\mathmakebox[.7cm][r]{16.343}\\{\mathmakebox[.7cm][r]{\scriptscriptstyle \pm 0.0008}}}$\\
\addlinespace
return\_simple&$\substack{\mathmakebox[.7cm][r]{27.678}\\{\mathmakebox[.7cm][r]{\scriptscriptstyle \pm 0.0273}}}$&$\substack{\mathmakebox[.7cm][r]{1.767}\\{\mathmakebox[.7cm][r]{\scriptscriptstyle \pm 0.0005}}}$&$\substack{\mathmakebox[.7cm][r]{0.251}\\{\mathmakebox[.7cm][r]{\scriptscriptstyle \pm 0.0000}}}$&$\substack{\mathmakebox[.7cm][r]{0.251}\\{\mathmakebox[.7cm][r]{\scriptscriptstyle \pm 0.0000}}}$&$\substack{\mathmakebox[.7cm][r]{0.195}\\{\mathmakebox[.7cm][r]{\scriptscriptstyle \pm 0.0000}}}$&$\substack{\mathmakebox[.7cm][r]{0.251}\\{\mathmakebox[.7cm][r]{\scriptscriptstyle \pm 0.0000}}}$&$\substack{\mathmakebox[.7cm][r]{0.223}\\{\mathmakebox[.7cm][r]{\scriptscriptstyle \pm 0.0000}}}$&$\substack{\mathmakebox[.7cm][r]{21.239}\\{\mathmakebox[.7cm][r]{\scriptscriptstyle \pm 0.0161}}}$\\
\addlinespace
scopes&$\substack{\mathmakebox[.7cm][r]{17.854}\\{\mathmakebox[.7cm][r]{\scriptscriptstyle \pm 0.0065}}}$&$\substack{\mathmakebox[.7cm][r]{2.009}\\{\mathmakebox[.7cm][r]{\scriptscriptstyle \pm 0.0002}}}$&$\substack{\mathmakebox[.7cm][r]{0.603}\\{\mathmakebox[.7cm][r]{\scriptscriptstyle \pm 0.0003}}}$&$\substack{\mathmakebox[.7cm][r]{0.134}\\{\mathmakebox[.7cm][r]{\scriptscriptstyle \pm 0.0000}}}$&$\substack{\mathmakebox[.7cm][r]{0.124}\\{\mathmakebox[.7cm][r]{\scriptscriptstyle \pm 0.0001}}}$&$\substack{\mathmakebox[.7cm][r]{0.601}\\{\mathmakebox[.7cm][r]{\scriptscriptstyle \pm 0.0002}}}$&$\substack{\mathmakebox[.7cm][r]{0.134}\\{\mathmakebox[.7cm][r]{\scriptscriptstyle \pm 0.0000}}}$&$\substack{\mathmakebox[.7cm][r]{20.412}\\{\mathmakebox[.7cm][r]{\scriptscriptstyle \pm 0.0014}}}$\\
\addlinespace
smallfunc&$\substack{\mathmakebox[.7cm][r]{46.912}\\{\mathmakebox[.7cm][r]{\scriptscriptstyle \pm 0.0368}}}$&$\substack{\mathmakebox[.7cm][r]{3.278}\\{\mathmakebox[.7cm][r]{\scriptscriptstyle \pm 0.0002}}}$&$\substack{\mathmakebox[.7cm][r]{0.251}\\{\mathmakebox[.7cm][r]{\scriptscriptstyle \pm 0.0000}}}$&$\substack{\mathmakebox[.7cm][r]{0.251}\\{\mathmakebox[.7cm][r]{\scriptscriptstyle \pm 0.0000}}}$&$\substack{\mathmakebox[.7cm][r]{0.188}\\{\mathmakebox[.7cm][r]{\scriptscriptstyle \pm 0.0000}}}$&$\substack{\mathmakebox[.7cm][r]{0.251}\\{\mathmakebox[.7cm][r]{\scriptscriptstyle \pm 0.0000}}}$&$\substack{\mathmakebox[.7cm][r]{0.251}\\{\mathmakebox[.7cm][r]{\scriptscriptstyle \pm 0.0000}}}$&$\substack{\mathmakebox[.7cm][r]{57.862}\\{\mathmakebox[.7cm][r]{\scriptscriptstyle \pm 0.0025}}}$\\
\addlinespace
sum&$\substack{\mathmakebox[.7cm][r]{23.612}\\{\mathmakebox[.7cm][r]{\scriptscriptstyle \pm 0.0201}}}$&$\substack{\mathmakebox[.7cm][r]{1.440}\\{\mathmakebox[.7cm][r]{\scriptscriptstyle \pm 0.0000}}}$&$\substack{\mathmakebox[.7cm][r]{0.074}\\{\mathmakebox[.7cm][r]{\scriptscriptstyle \pm 0.0000}}}$&$\substack{\mathmakebox[.7cm][r]{0.074}\\{\mathmakebox[.7cm][r]{\scriptscriptstyle \pm 0.0000}}}$&$\substack{\mathmakebox[.7cm][r]{0.056}\\{\mathmakebox[.7cm][r]{\scriptscriptstyle \pm 0.0000}}}$&$\substack{\mathmakebox[.7cm][r]{0.074}\\{\mathmakebox[.7cm][r]{\scriptscriptstyle \pm 0.0000}}}$&$\substack{\mathmakebox[.7cm][r]{0.065}\\{\mathmakebox[.7cm][r]{\scriptscriptstyle \pm 0.0000}}}$&$\substack{\mathmakebox[.7cm][r]{31.124}\\{\mathmakebox[.7cm][r]{\scriptscriptstyle \pm 0.0063}}}$\\
\addlinespace
sum\_meth&$\substack{\mathmakebox[.7cm][r]{25.428}\\{\mathmakebox[.7cm][r]{\scriptscriptstyle \pm 0.0994}}}$&$\substack{\mathmakebox[.7cm][r]{1.793}\\{\mathmakebox[.7cm][r]{\scriptscriptstyle \pm 0.0021}}}$&$\substack{\mathmakebox[.7cm][r]{0.074}\\{\mathmakebox[.7cm][r]{\scriptscriptstyle \pm 0.0000}}}$&$\substack{\mathmakebox[.7cm][r]{0.074}\\{\mathmakebox[.7cm][r]{\scriptscriptstyle \pm 0.0000}}}$&&$\substack{\mathmakebox[.7cm][r]{0.074}\\{\mathmakebox[.7cm][r]{\scriptscriptstyle \pm 0.0000}}}$&$\substack{\mathmakebox[.7cm][r]{0.065}\\{\mathmakebox[.7cm][r]{\scriptscriptstyle \pm 0.0000}}}$&$\substack{\mathmakebox[.7cm][r]{33.283}\\{\mathmakebox[.7cm][r]{\scriptscriptstyle \pm 0.0360}}}$\\
\addlinespace
sum\_meth\_attr&$\substack{\mathmakebox[.7cm][r]{31.930}\\{\mathmakebox[.7cm][r]{\scriptscriptstyle \pm 0.0046}}}$&$\substack{\mathmakebox[.7cm][r]{4.351}\\{\mathmakebox[.7cm][r]{\scriptscriptstyle \pm 0.0003}}}$&$\substack{\mathmakebox[.7cm][r]{0.243}\\{\mathmakebox[.7cm][r]{\scriptscriptstyle \pm 0.0003}}}$&$\substack{\mathmakebox[.7cm][r]{0.243}\\{\mathmakebox[.7cm][r]{\scriptscriptstyle \pm 0.0014}}}$&&$\substack{\mathmakebox[.7cm][r]{0.275}\\{\mathmakebox[.7cm][r]{\scriptscriptstyle \pm 0.0001}}}$&$\substack{\mathmakebox[.7cm][r]{0.220}\\{\mathmakebox[.7cm][r]{\scriptscriptstyle \pm 0.0003}}}$&$\substack{\mathmakebox[.7cm][r]{35.305}\\{\mathmakebox[.7cm][r]{\scriptscriptstyle \pm 0.0125}}}$\\
\addlinespace
total\_list&$\substack{\mathmakebox[.7cm][r]{8.076}\\{\mathmakebox[.7cm][r]{\scriptscriptstyle \pm 0.0058}}}$&$\substack{\mathmakebox[.7cm][r]{0.943}\\{\mathmakebox[.7cm][r]{\scriptscriptstyle \pm 0.0003}}}$&$\substack{\mathmakebox[.7cm][r]{0.363}\\{\mathmakebox[.7cm][r]{\scriptscriptstyle \pm 0.0001}}}$&$\substack{\mathmakebox[.7cm][r]{0.420}\\{\mathmakebox[.7cm][r]{\scriptscriptstyle \pm 0.0001}}}$&$\substack{\mathmakebox[.7cm][r]{0.633}\\{\mathmakebox[.7cm][r]{\scriptscriptstyle \pm 0.0001}}}$&$\substack{\mathmakebox[.7cm][r]{0.360}\\{\mathmakebox[.7cm][r]{\scriptscriptstyle \pm 0.0002}}}$&$\substack{\mathmakebox[.7cm][r]{0.246}\\{\mathmakebox[.7cm][r]{\scriptscriptstyle \pm 0.0001}}}$&$\substack{\mathmakebox[.7cm][r]{14.138}\\{\mathmakebox[.7cm][r]{\scriptscriptstyle \pm 0.0257}}}$\\
\addlinespace
walk\_list&$\substack{\mathmakebox[.7cm][r]{1.054}\\{\mathmakebox[.7cm][r]{\scriptscriptstyle \pm 0.0007}}}$&$\substack{\mathmakebox[.7cm][r]{0.085}\\{\mathmakebox[.7cm][r]{\scriptscriptstyle \pm 0.0000}}}$&$\substack{\mathmakebox[.7cm][r]{0.162}\\{\mathmakebox[.7cm][r]{\scriptscriptstyle \pm 0.0001}}}$&$\substack{\mathmakebox[.7cm][r]{0.208}\\{\mathmakebox[.7cm][r]{\scriptscriptstyle \pm 0.0003}}}$&$\substack{\mathmakebox[.7cm][r]{0.333}\\{\mathmakebox[.7cm][r]{\scriptscriptstyle \pm 0.0003}}}$&$\substack{\mathmakebox[.7cm][r]{0.210}\\{\mathmakebox[.7cm][r]{\scriptscriptstyle \pm 0.0003}}}$&$\substack{\mathmakebox[.7cm][r]{0.225}\\{\mathmakebox[.7cm][r]{\scriptscriptstyle \pm 0.0001}}}$&$\substack{\mathmakebox[.7cm][r]{2.218}\\{\mathmakebox[.7cm][r]{\scriptscriptstyle \pm 0.0144}}}$\\
\midrule
deltablue&$\substack{\mathmakebox[.7cm][r]{0.901}\\{\mathmakebox[.7cm][r]{\scriptscriptstyle \pm 0.0006}}}$&$\substack{\mathmakebox[.7cm][r]{36.609}\\{\mathmakebox[.7cm][r]{\scriptscriptstyle \pm 0.0320}}}$&$\substack{\mathmakebox[.7cm][r]{0.236}\\{\mathmakebox[.7cm][r]{\scriptscriptstyle \pm 0.0005}}}$&$\substack{\mathmakebox[.7cm][r]{0.055}\\{\mathmakebox[.7cm][r]{\scriptscriptstyle \pm 0.0002}}}$&&$\substack{\mathmakebox[.7cm][r]{0.246}\\{\mathmakebox[.7cm][r]{\scriptscriptstyle \pm 0.0005}}}$&$\substack{\mathmakebox[.7cm][r]{0.025}\\{\mathmakebox[.7cm][r]{\scriptscriptstyle \pm 0.0001}}}$&$\substack{\mathmakebox[.7cm][r]{7.860}\\{\mathmakebox[.7cm][r]{\scriptscriptstyle \pm 0.1417}}}$\\
\addlinespace
fannkuch&$\substack{\mathmakebox[.7cm][r]{15.273}\\{\mathmakebox[.7cm][r]{\scriptscriptstyle \pm 0.0168}}}$&$\substack{\mathmakebox[.7cm][r]{2.480}\\{\mathmakebox[.7cm][r]{\scriptscriptstyle \pm 0.0017}}}$&$\substack{\mathmakebox[.7cm][r]{1.371}\\{\mathmakebox[.7cm][r]{\scriptscriptstyle \pm 0.0004}}}$&$\substack{\mathmakebox[.7cm][r]{0.742}\\{\mathmakebox[.7cm][r]{\scriptscriptstyle \pm 0.0002}}}$&$\substack{\mathmakebox[.7cm][r]{1.403}\\{\mathmakebox[.7cm][r]{\scriptscriptstyle \pm 0.0002}}}$&$\substack{\mathmakebox[.7cm][r]{1.393}\\{\mathmakebox[.7cm][r]{\scriptscriptstyle \pm 0.0002}}}$&$\substack{\mathmakebox[.7cm][r]{0.746}\\{\mathmakebox[.7cm][r]{\scriptscriptstyle \pm 0.0002}}}$&$\substack{\mathmakebox[.7cm][r]{10.676}\\{\mathmakebox[.7cm][r]{\scriptscriptstyle \pm 0.0092}}}$\\
\addlinespace
mandel&&$\substack{\mathmakebox[.7cm][r]{0.460}\\{\mathmakebox[.7cm][r]{\scriptscriptstyle \pm 0.0033}}}$&$\substack{\mathmakebox[.7cm][r]{0.536}\\{\mathmakebox[.7cm][r]{\scriptscriptstyle \pm 0.0003}}}$&$\substack{\mathmakebox[.7cm][r]{0.581}\\{\mathmakebox[.7cm][r]{\scriptscriptstyle \pm 0.0002}}}$&$\substack{\mathmakebox[.7cm][r]{0.287}\\{\mathmakebox[.7cm][r]{\scriptscriptstyle \pm 0.0000}}}$&$\substack{\mathmakebox[.7cm][r]{0.581}\\{\mathmakebox[.7cm][r]{\scriptscriptstyle \pm 0.0001}}}$&&$\substack{\mathmakebox[.7cm][r]{4.211}\\{\mathmakebox[.7cm][r]{\scriptscriptstyle \pm 0.0112}}}$\\
\addlinespace
richards&$\substack{\mathmakebox[.7cm][r]{11.901}\\{\mathmakebox[.7cm][r]{\scriptscriptstyle \pm 0.0055}}}$&$\substack{\mathmakebox[.7cm][r]{5.263}\\{\mathmakebox[.7cm][r]{\scriptscriptstyle \pm 0.0027}}}$&$\substack{\mathmakebox[.7cm][r]{0.377}\\{\mathmakebox[.7cm][r]{\scriptscriptstyle \pm 0.0004}}}$&$\substack{\mathmakebox[.7cm][r]{0.442}\\{\mathmakebox[.7cm][r]{\scriptscriptstyle \pm 0.0002}}}$&&$\substack{\mathmakebox[.7cm][r]{0.392}\\{\mathmakebox[.7cm][r]{\scriptscriptstyle \pm 0.0003}}}$&$\substack{\mathmakebox[.7cm][r]{0.216}\\{\mathmakebox[.7cm][r]{\scriptscriptstyle \pm 0.0002}}}$&$\substack{\mathmakebox[.7cm][r]{10.709}\\{\mathmakebox[.7cm][r]{\scriptscriptstyle \pm 0.0091}}}$\\
\bottomrule
\end{tabular}
\end{adjustbox}
\caption{Absolute microbenchmark timings.}
\label{tab:absresults}
\end{table*}


\end{document}